\documentclass[twocolumn]{aastex63}

\defcitealias{2019ApJCambioni}{C19}
\defcitealias{EmsenhuberApj2020}{E20}
\def\paperone{\citetalias{2019ApJCambioni}}
\def\papertwo{\citetalias{EmsenhuberApj2020}}

\def\sci{Science}
\def\epsl{EPSL}
\def\natgeo{NatGeo}

\usepackage{savesym}
\savesymbol{tablenum}
\usepackage{multirow}
\usepackage{siunitx}
\restoresymbol{SIX}{tablenum}

\usepackage{amsmath}
\usepackage{amssymb}
\usepackage{wasysym}
\usepackage{graphicx}
\usepackage{xcolor}
\usepackage{natbib}	
\usepackage{enumitem}

\DeclareSIUnit\au{au}
\DeclareSIUnit\mearth{M_\oplus}
\DeclareSIUnit\year{yr}

\begin{document}

    \title{The Effect of Inefficient Accretion on Planetary Differentiation}

\correspondingauthor{Saverio Cambioni}
\email{cambioni@lpl.arizona.edu}

\author[0000-0001-6294-4523]{Saverio Cambioni}
\affil{Lunar and Planetary Laboratory, University of Arizona, 1629 E.\ University Blvd., Tucson, AZ 85721, USA}

\author[0000-0002-4952-9007]{Seth A. Jacobson}
\affil{Department of Earth and Environmental Sciences, Michigan State University, 288 Farm Ln, East Lansing, MI 48824, USA}

\author[0000-0002-8811-1914]{Alexandre Emsenhuber}
\affil{Lunar and Planetary Laboratory, University of Arizona, 1629 E.\ University Blvd., Tucson, AZ 85721, USA}

\author[0000-0003-1002-2038]{Erik Asphaug}
\affil{Lunar and Planetary Laboratory, University of Arizona, 1629 E.\ University Blvd., Tucson, AZ 85721, USA}

\author[0000-0002-3009-1017]{David C. Rubie}
\affil{Bayerisches Geoinstitut, University of Bayreuth, D-95440 Bayreuth, Germany}

\author[0000-0002-9767-4153]{Travis S. J. Gabriel}
\affil{School of Earth and Space Exploration, Arizona State University, 781 E.\ Terrace Mall, Tempe, AZ 85287, USA}

\author[0000-0001-5475-9379]{Stephen R. Schwartz}
\affil{Lunar and Planetary Laboratory, University of Arizona, 1629 E.\ University Blvd., Tucson, AZ 85721, USA}

\author[0000-0001-6076-8992]{Roberto Furfaro}
\affil{Department of Systems and Industrial Engineering,\ Department of Aerospace and Mechanical Engineering, University of Arizona,\ 1127 E. James E. Rogers Way, Tucson, AZ 85721, USA}

\begin{abstract}
Pairwise collisions between terrestrial embryos are the dominant means of accretion during the last stage of planet formation. Hence, their realistic treatment in \textit{N}-body studies is critical to accurately model the formation of terrestrial planets and to develop interpretations of telescopic and spacecraft observations. In this work, we compare the effects of two collision prescriptions on the core-mantle differentiation of terrestrial planets: a model in which collisions are always completely accretionary (``perfect merging'') and a more realistic model based on neural networks that has been trained on hydrodynamical simulations of giant impacts. The latter model is able to predict the loss of mass due to imperfect accretion and the evolution of non-accreted projectiles in hit-and-run collisions. We find that the results of the neural-network model feature a wider range of final core mass fractions and metal-silicate equilibration pressures, temperatures, and oxygen fugacities than the assumption of perfect merging. When used to model collisions in \textit{N}-body studies of terrestrial planet formation, the two models provide similar answers for planets more massive than \SI{0.1}{\mearth} (Earth's masses). For less massive final bodies, however, the inefficient-accretion model predicts a higher degree of compositional diversity. This phenomenon is not reflected in planet formation models of the solar system that use perfect merging to determine collisional outcomes. Our findings confirm the role of giant impacts as important drivers of planetary diversity and encourage a realistic implementation of inefficient accretion in future accretion studies.
\end{abstract}

\keywords{Planetary interior (1248) --- Planetary system formation (1257) --- Inner planets (797)}

\section{Introduction}
\label{sec:intro}

Collisions between similar-size planetary bodies (``giant impacts'') dominate the final stage of planet formation \citep{1985ScienceWetherill,2010ChEGAsphaug}.
These events generally result in the formation of transient magma oceans on the resulting bodies \citep{1993TonksMelosh,2016PEPSdeVries}. During the last decade, a series of studies combined core-mantle differentiation with accretion modeling to put constraints on how terrestrial planets' cores formed in the solar system and showed that core formation of terrestrial planets does not occur in a single stage, but rather that it is the result of a multistage process, i.e., a series of metal-silicate equilibrations  \citep[e.g.,][]{2011Rubie,Rubie2015,2015IcarDwyer,2015IcarBonsor,Carter2015apj,2016Rubie,2017EPSLFischer,2019EPSLZube}. In particular, the core formation model from \citet{Rubie2015} uses rigorous chemical mass balance with metal-silicate element partitioning data and requires assumptions regarding the bulk compositions of all starting embryos and planetesimals as a function of heliocentric distance.
The differentiation of terrestrial planets is modeled as the separation of the iron-rich metal from silicate material using metal-silicate partitioning of elements to determine the evolving chemistry of the two reservoirs. New insights into terrestrial planet formation have been enabled by this equilibration model. For example, \citet{Rubie2015} demonstrated that Earth likely accreted from heterogeneous reservoirs of early solar system materials, \citet{2016Rubie} demonstrated that iron sulfide segregation was responsible for stripping the highly siderophile elements from Earth’s mantle, and \citet{2017JacobsonEPSL} proposed that Venus does not have an active planetary dynamo because it never experienced late Moon-forming giant impacts.

These previous studies of planet differentiation interpreted the results of \textit{N}-body simulations of terrestrial planet formation where collisions were treated as perfectly inelastic, which computer models of giant impacts have shown is an oversimplification of the complex collision process \citep[e.g.,][]{2004ApJAgnor,2012ApJ...751...32S}. Inelastic collisions, or ``perfect merging", assumes the projectile mass ($M_\mathrm{P}$) merges with the target mass ($M_\mathrm{T}$) to form a body with mass $M_\mathrm{T}+ M_\mathrm{P}$. However, in nearly all giant impacts, escaping debris is produced, and the projectile's core does not simply descend through the magma ocean and merge with the target's metal core.
Instead, half of all collisions are `hit-and-run', where the projectile escapes accretion \citep[][]{2004ApJAgnor,2010ApJKokubo} and may never re-impact the target again \citep{2013IcarusChambers,EmsenhuberApj2020}.
In these events, partial accretion may occur between the metal and silicate reservoirs of the target body and the projectile (or `runner'). To accurately model the geochemical evolution of the mantles and cores of growing planetary bodies, it is thus necessary to account for the range of accretionary (or non-accretionary) outcomes of giant impacts.

\subsection{Beyond perfect merging}

High-resolution hydrocode computer simulations of collisions provide a description of the outcomes of giant impacts, which can then be incorporated into planet formation models to produce higher-fidelity predictions. But each giant impact simulation requires a long computational time to complete (on the order of hours to days depending on the resolution and computing resources).
Since a large number of collisions may occur during late-stage terrestrial planet formation \citep[up to order of \num{e3}, e.g.,][]{EmsenhuberApj2020}, it is impractical to model each impact ``on-the-fly'' by running a full hydrocode simulation at a resolution that is sufficient to make meaningful predictions.

Several previous studies focused on overcoming both the assumption of perfect merging and the aforementioned computational bottleneck. These works employed various techniques to resolve different aspects of giant impacts. Commonly, scaling laws and other algebraic relationships \citep[e.g.][]{1986MmSAI..57...65H,HOUSEN1990226,HOLSAPPLE19941067,2010ApJKokubo,2012ApJLeinhardt,2017IcarusGenda} are utilized during an \textit{N}-body (orbital dynamical) planet formation simulation to predict the collision outcome for the masses of post-impact remnants \citep[e.g.][]{2013IcarusChambers,2016ApJQuintana,2019AJClementA}. The post-impact information is then fed back into the \textit{N}-body code for further dynamical evolution. Other studies `handed off' each collision scenario to a hydrodynamic simulation in order to model the exact impact scenario explicitly, and fed post-impact information back to the \textit{N}-body code \citep{2017IcarusGenda,2020A&ABurger}. The latter methodology is the most rigourous, but also the most computationally demanding, as it requires to run a hydrodynamic calculation for every one of the hundreds of collisions in
an \textit{N}-body simulation, each of which requires days of computer time when using modest resolution.

Alternatively, \citet[termed \paperone{} hereafter]{2019ApJCambioni} proposed a fully data-driven approach, in which machine-learning algorithms were trained on a data set of pre-existing hydrocode simulations \citep{2011PhDReufer,2020ApJGabriel}.
The machine-learned functions (``surrogate models") predict the outcome of a collision within a known level of accuracy with respect to the hydrocode simulations in an independent testing set. This process is fully data-driven and does not introduce model assumptions in the fitting, which is in contrast to scaling laws composed of a set of algebraic functions based on physical arguments \citep[e.g.,][]{2012ApJLeinhardt,2020ApJGabriel}. The surrogate models are fast predictors and \citet[][termed \papertwo~hereafter]{EmsenhuberApj2020} implemented them in a code library named \texttt{collresolve}\footnote{\url{https://github.com/aemsenhuber/collresolve}} \citep{2019SoftwareEmsenhuberCambioni} to realistically treat collisions on-the-fly during terrestrial planet formation studies. 
When \texttt{collresolve} is used to treat collisions in \textit{N}-body studies, the final planets feature a wider range of masses and degree of mixing across feeding zones in the disk compared with those predicted by assuming perfect merging. Although \papertwo~ignored debris re-accretion ---and we use these dynamical simulations for the study herein--- their results suggest that composition diversity increases in collision remnants. This is something that cannot be predicted by models that assume perfect merging.

\subsection{This work}

In this paper, we compare the collision outcome obtained assuming perfect merging with that predicted by the more realistic machine-learned giant impact model of \paperone~and \papertwo{}.
In the former case, debris is not produced by definition and in the latter case, debris is produced but not re-accreted.
In this respect, our goal is not to reproduce the solar system terrestrial planets, but to investigate whether or not the two collision models produce different predictions in terms of terrestrial planets' core-mantle differentiation at the end of the planetary system's dynamical evolution.

\paperone~and~\papertwo~developed models for the mass and orbits of the largest post-impact remnants; here we go a step further and develop a model for the preferential erosion of mantle silicates and core materials. To do so, we train two new neural networks to predict the core mass fraction of the resulting bodies of a giant impact.
We describe the data-driven model of inefficient accretion by \paperone~and~\papertwo{} in Section \ref{sec:ML} and its implementation in the core-mantle differentiation model by \citet{Rubie2015} in Section \ref{sec:equi_method}.

We compare the perfect merging and inefficient-accretion models in two ways: (1) by studying the case of a single collision between two planetary embryos (Section \ref{sec:map_elements}); and (2) by interpreting the effect of multiple giant impacts in the \textit{N}-body simulations of accretion presented in \papertwo~(Section \ref{sec:summary_N_body}).
During the accretion of planets through giant impacts between planetary embryos, we focus on the evolution of those variables that control planetary differentiation: mass, core mass fraction, as well as metal-silicate equilibration pressure, temperature, and oxygen fugacity. Other factors that may alter composition and thermodynamical evolution indirectly, e.g., atmospheric escape and radiative effects, are not covered in these models. 

\section{Inefficient-accretion model}
\label{sec:ML}
The data-driven inefficient-accretion model by \paperone{} and \papertwo{} consists of applying machine learning to the prediction of giant impacts' outcomes based on the pre-existing set of collision simulations described below in Section \ref{sec:SPH_data}.
By training on a large data set of simulations of giant impacts, this approach allows producing response functions (surrogate models) that accurately and quickly predict key outcomes of giant impacts needed to introduce realistic collision outcomes ``on-the-fly" of an \textit{N}-body code. 

\subsection{Data set of giant impact simulations}
\label{sec:SPH_data}

The data set used in \paperone~and~\papertwo~and in this work is composed of nearly 800 simulations of planetary collisions performed using the Smoothed-Particle Hydrodynamics (SPH) technique \citep[see, e.g.,][for reviews]{1992ARA&AMonaghan,2009NARRosswog} obtained by \citet{2011PhDReufer} and further described in \citet{2020ApJGabriel}.
They have a resolution of $\sim$ \num{2e5} SPH particles.
All bodies are differentiated with a bulk composition of 70 wt\% silicate and 30 wt\% metallic iron, where the equation of state for iron is ANEOS \citep{ANEOS} and M-ANEOS for SiO\textsubscript{2} \citep{2007M&PSMelosh}.
The data set spans target masses $M_\mathrm{T}$ from \num{e-2} to \SI{1}{\mearth}, projectile-to-target mass ratios $\gamma=M_\mathrm{P}/M_\mathrm{T}$ between \num{0.2} and \num{0.7}, all impact angles $\theta_{coll}$, and impact velocities $v_\mathrm{coll}$ between 1 and 4 times the mutual escape velocity $v_\mathrm{esc}$, where
\begin{equation}
\label{V_esc}
v_\mathrm{esc}=\sqrt{\frac{2G(M_\mathrm{T}+M_\mathrm{P})}{R_\mathrm{T}+R_\mathrm{P}}},
\end{equation}
\noindent
which represents the entire range of expected impact velocities between major bodies from \textit{N}-body models \citep[e.g.][]{2013IcarusChambers,2016ApJQuintana}.
In Equation \ref{V_esc}, $G$ is the gravitational constant, and $R_\mathrm{T}$ and $R_\mathrm{P}$ are the bodies' radii.
We refer to \paperone, \papertwo, and \citet{2020ApJGabriel} for more information about the data set. An excerpt of the data set is reported in Table \ref{tab:data}.
The data set is provided in its entirety in the machine-readable format.
\textbf{}
 \begin{table*}
    \centering
    \caption{Excerpt of the data from the collision simulation analysis.}
    \begin{tabular}{cccc|c|cccc}
        \hline
        Target mass & Mass ratio & Angle & Velocity & Type & Acc. L. R. & Acc. S. R. & CMF L. R.  & CMF S. R. \\
        $M_\mathrm{T}\ [\si{\mearth}]$ & $\gamma=M_\mathrm{P}/M_\mathrm{T}$ & $\theta_{coll}$ [deg] & $v_\mathrm{coll}/v_\mathrm{esc}$ & & $\xi_\mathrm{L}$ & $\xi_\mathrm{S}$ & $Z_\mathrm{L}$ & $Z_\mathrm{S}$ \\
        \hline
        \hline
        \num{1}   & 0.70 & 52.5 & 1.15 &  1 &  0.02 & -0.03 & 0.30 & 0.31 \\
        \num{1}   & 0.70 & 22.5 & 3.00 &  1 & -0.58 & -0.62 & 0.50  &  0.62  \\
        \num{1}   & 0.70 & 45.0 & 1.30 &  1 &  0.02 & -0.04 & 0.30 & 0.31\\
        \num{e-1} & 0.70 & 15.0 & 1.40 &  0 &  0.90 & -1.00 & 0.31 & ...  \\
        \num{e-1} & 0.20 & 15.0 & 3.50 & -1 & -1.51 & -1.00 & 0.43 & ...  \\
        \num{e-1} & 0.35 & 15.0 & 3.50 & -1 & -1.25 & -1.00 & 0.50 & ...  \\
        \num{e-2} & 0.70 & 60.0 & 1.70 &  1 &  0.00 & -0.02 & 0.30 &   0.30  \\
        \hline
    \end{tabular}
    \tablecomments{This table is published in its entirety in the machine-readable format. A portion is shown here for guidance regarding its form and content.}
    \tablecomments{The elements in the first four columns are the predictors (collision properties): $M_T \in [\num{e-2},~1]~\si{\mearth}$; $\gamma=M_\mathrm{P}/M_\mathrm{T} \in [0.2,~0.7]$; $\theta_{coll} \in [0,~\ang{90}]$; $v_{coll}/v_{esc} \in [1,~4]$. The elements in the other columns are the responses describing the collision outcome. The fifth column (Type) is the automated classification (for the classifier), with hit-and-run collisions coded as a 1, accretion as a 0, and erosion as a -1. The last four columns are the responses for the neural networks: $\xi_\mathrm{L}$ is computed according to Equation~(\ref{eq:acclr}), $\xi_\mathrm{S}$ to Equation~(\ref{eq:accsr}), and the post-collision Core Mass Fractions (CMF) of the largest remnant and second remnant, $Z_\mathrm{L}$ and $Z_\mathrm{S}$ respectively, are defined in Section \ref{sec:regressor_CMF}.}
    \label{tab:data}
\end{table*}

Designing the surrogate models described in the following Sections requires running hundreds of SPH simulations and training the machine-learning functions. The computational cost of this procedure, however, is low when compared against the computational resources necessary to solve each collision in a \textit{N}-body study with a full SPH simulation \added{at the same particle resolution of the simulations used in \paperone{} and \papertwo} for each event instead of using the surrogate models. Each giant impact simulation requires a long computational time to complete, on the order of hours to days depending on the resolution and computing resources, while the surrogate models, once constructed, provide an answer in a fraction of a second (\paperone;~\papertwo).

\subsection{Surrogate model of accretion efficiencies}
\label{sec:regressor_accs}
In order to assess the accretion efficiency of the target across simulations of varying masses of targets and projectiles, \papertwo~normalize the change in mass of the largest remnant, assumed to be the post-impact target, by the projectile mass \citep{2010ChEGAsphaug}:
\begin{equation}
    \xi_\mathrm{L}=\frac{M_\mathrm{L}-M_\mathrm{T}}{M_\mathrm{P}} < 1,
    \label{eq:acclr}
\end{equation}
\noindent
where $M_\mathrm{L}$ is the mass of the largest single gravitationally bound remnant.
Accretion onto the target causes $\xi_\mathrm{L}>0$, while negative values indicate erosion.
This accretion efficiency is heavily dependent on the impact velocity relative to the mutual escape velocity $v_\mathrm{esc}$, especially in the critical range $\sim1$--$1.4~v_\mathrm{esc}$, which encompasses $\sim$90\% of the probability distribution of impact velocities between major remnants \added{in the $N$-body simulations by \citet{2013IcarusChambers} which include collision fragmentation \citep{2020ApJGabriel}.}

Similarly, for the second largest remnant with mass $M_\mathrm{S}$, assuming it is the post-impact projectile, i.e., the runner in a hit-and-run collision \citep{2006NatureAsphaug}, \papertwo~define a non-dimensional accretion efficiency again normalized by the projectile mass:
\begin{equation}
    \xi_\mathrm{S}=\frac{M_\mathrm{S} - M_\mathrm{P}}{M_\mathrm{P}}.
    \label{eq:accsr}
\end{equation}
The value $\xi_\mathrm{S}$ is almost always negative, as mass transfer from the projectile onto the target occurs also in case of projectile survival \citep{2006NatureAsphaug,2019ApJEmsenhuberA}, and loss to debris can occur.

The mass of the debris $M_\mathrm{D}$ is computed from mass conservation.
If the debris creation efficiency is defined as:  $\xi_\mathrm{D}=M_\mathrm{D}/M_\mathrm{P}$, then $\xi_\mathrm{L}+\xi_\mathrm{S}+\xi_\mathrm{D}=0$. 

In \papertwo, the quantities of Equations \ref{eq:acclr} and \ref{eq:accsr} were used to train a surrogate model of accretion efficiencies.
This is a neural network, that is, a parametric function trained to mimic the ``parent'' SPH calculation as an input-output function, in order to predict real-variable outputs given the four impact parameters (predictors): mass of the target, projectile-to-target mass ratio, impact angle, and impact velocity.
The data set entries are of the type: 
\begin{equation}
    \{(M_\mathrm{T},~\gamma,~\theta_\mathrm{coll},~v_\mathrm{coll}/v_\mathrm{esc}) ; (\xi_\mathrm{L}, \xi_\mathrm{S})\}.
\end{equation}

The surrogate model is assessed in its training success and predictive capabilities by means of the mean squared error
\begin{equation}
\label{eq:MSE}
    \frac{1}{N} \sum_{i=0}^N (Q_\mathrm{NN}-Q_\mathrm{SPH})^2,
\end{equation}
\noindent
and correlation coefficient
\begin{equation}
\label{eq:R-value}
    \frac{\mathrm{cov}(Q_\mathrm{NN},Q_\mathrm{SPH})}{\sigma_\mathrm{NN} \sigma_\mathrm{SPH}}
\end{equation}
for each quantity $Q$, where $Q_\mathrm{NN}$ and $Q_\mathrm{SPH}$ indicate the predictions by the Neural Networks and the correspondent outcome from the SPH simulations with standard deviations $\sigma_\mathrm{NN}$ and $\sigma_\mathrm{SPH}$, respectively. The goal is to achieve a mean squared error as close to zero and a correlation coefficient as close to 100\% as possible on a testing data set, which comprises data that were not used for training. The surrogate model of accretion efficiency is able to predict the mass of the largest and second largest remnants with a mean squared error at testing equal to 0.03 and a correlation coefficient greater than 96\%.

Importantly, although the surrogate model has a high global accuracy, inaccurate predictions can still occur locally in the parameter space (\paperone). \added{These inaccurate predictions, however, are not systematic; the distributions of the residuals $\Delta \xi$ between accretion efficiency predictions $Q_{NN}$ and target values $Q_{SPH}$ for the largest remnant and second remnant are well fit by Gaussian distributions $\mathcal{N} (\mu, \sigma) = \mathcal{N} (0.0,0.1)$ and $\mathcal{N} (0.02,0.09)$, respectively, where $\mu$ is the mean of the residuals and $\sigma$ is the standard deviation of the residuals, so that $\mathcal{N} (0, 0)$ is the distribution associated with a noiseless surrogate model. The noise level is comparable to the numerical precision of the SPH simulations ($\sim$ 0.1--0.15 in units of accretion efficiency). High-inaccuracy predictions (i.e., $|\Delta \xi| > 0.5$) account for just 3.7\% and 2.6\% of the overall set for largest and second remnants respectively, and tend to cluster in proximity of the boundaries between different regimes, as previously discussed in \paperone{}.}

\subsection{Classifier of collision types}
\label{sec:class_reg}
In \papertwo, the data set of SPH simulations of Section \ref{sec:SPH_data} was also used to train a classifier which provides predictions of the type of collisions (classes, or responses) based on the following mass criterion: accretion ($M_\mathrm{L}>M_\mathrm{T}$ and $M_\mathrm{S}<0.1 M_\mathrm{P}$), erosion ($M_\mathrm{L}<M_\mathrm{T}$ and $M_\mathrm{S}<0.1 M_\mathrm{P}$), and hit-and-run collision ($M_\mathrm{S}>0.1 M_\mathrm{P}$).
The data set entries are of the type: 

\begin{equation}
    \{(M_\mathrm{T},~\gamma,~\theta_\mathrm{coll},~v_\mathrm{coll}/v_\mathrm{esc}) ; \mathrm{type}\}
\end{equation}

\added{As opposed to the prediction by the surrogate model of accretion efficiency described in Section \ref{sec:regressor_accs}, the prediction of the classifier is categorical in type and its accuracy is computed as the mean value of the correct classifications over the whole population at testing. This accuracy is equal to 95\% globally, 83.3\% on the ``erosion'' type, 91.7\% for the ``accretion'' type, and 98.0\% for the hit-and-run collision type.}

\subsection{Surrogate models of core mass fraction}
\label{sec:regressor_CMF}

\begin{figure*}
	\centering
	\includegraphics[width=1\linewidth]{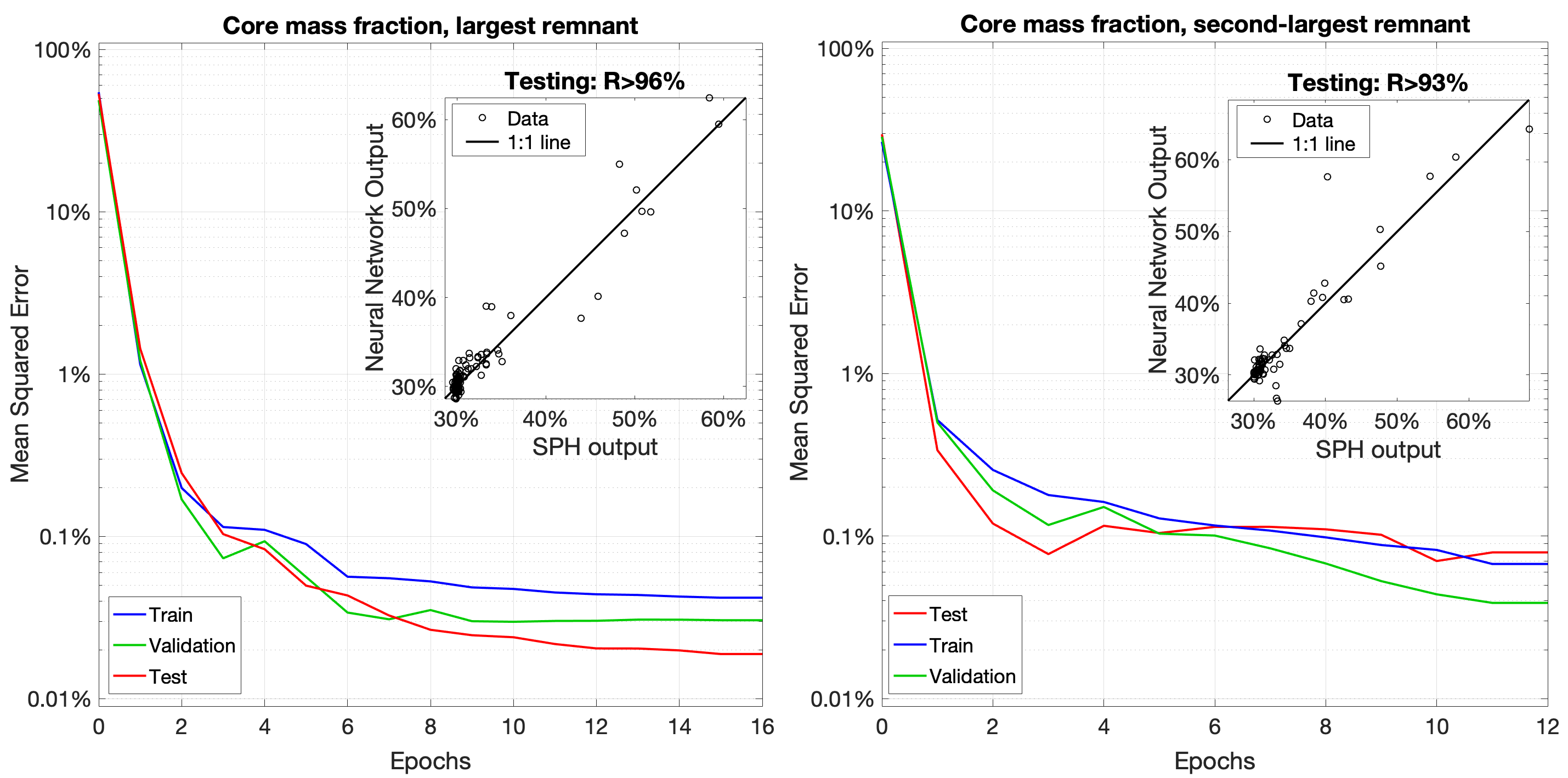}
	\caption{As the training proceeds, the mean squared errors (Equation \ref{eq:MSE}) of the surrogate models of core mass fraction on the training (blue curves), validation (green curves) and testing (red curves) data sets decrease at each training epoch until convergence is reached. The performances of the surrogate models are quantified in terms of mean squared error (Equation \ref{eq:MSE}) and correlation coefficient (Equation \ref{eq:R-value}) at testing: \{0.02\%; $R>96\%$\} and \{0.08\%; $R>93\%$\} for the largest and second largest remnants, respectively. The two inset plots show the correlation between predictions by the neural networks and the corresponding SPH values in the testing data sets (open dots) and the 1:1 correlation line.}
	\label{fig:reg_perf}
\end{figure*}

We use the same SPH data set as \paperone~and \papertwo{}~to train two new surrogate models to predict the core mass fractions of the largest and second largest remnants.
Each remnant's core mass fraction is obtained by accounting for all material in the SPH simulations.
This includes \emph{all} gravitationally-bound material such as potential silicate vapour resulting from energetic collisions involving larger bodies or impact velocities.
The core mass fractions of the target and projectile are termed $Z_\mathrm{T}$ and $Z_\mathrm{P}$, respectively. 
Their initial values are always equal to 30\% \added{in the dataset of SPH simulations used here}.

For the first new surrogate model, we train, validate, and test a neural network using a data set with entries:
\begin{equation}
    \{(M_\mathrm{T},~\gamma,~\theta_\mathrm{coll},~v_\mathrm{coll}/v_\mathrm{esc}) ; Z_\mathrm{L}\},
\end{equation}
where $Z_\mathrm{L}$ is the largest remnant's core mass fraction. In the hit-and-run regime only, we train a second neural network to predict the core mass fraction of the second largest remnant.
For this surrogate model, the data set has entries:
\begin{equation}
    \{(M_\mathrm{T},~\gamma,~\theta_\mathrm{coll},~v_\mathrm{coll}/v_\mathrm{esc}) ; Z_\mathrm{S}\},
\end{equation}
\noindent
where $Z_\mathrm{S}$ is the post-collision core mass fraction of the second largest remnant.

Following the approach described in \paperone{} and \papertwo{}, the training of the networks is performed on 70\% of the overall data set.
The rest of the data are split between a validation set (15\%) and a testing set (15\%) via random sampling without replacement. \added{Each neural network architecture consists of an input layer with 4 nodes (as many as the impact properties), one or more hidden layers and an output node. The number of hidden layers, number of neurons in each hidden layer, and the neurons activation functions (that is, the functions that define the output of the neurons given a set of inputs) are among the ``hyperparameters'' of the network. The optimal hyperparameters are not learned during training, but found through minimization of the mean squared error on the validation set (Equation \ref{eq:MSE}). Additional hyperparameters include the choice of the training algorithm, the intensity of the regularization of the training cost function aimed to avoid the choice for too ``complex'' models \citep[e.g., ][]{girosi1995regularization} and the strategy of data normalization before training.} The optimal neural network architectures have 10 neurons in the hidden layer with an hyperbolic tangent sigmoid activation function (Equation 13 in \paperone).
The inputs and targets are normalized in the range [-1, 1]. The regularization process \citep[e.g., ][]{girosi1995regularization} has a strength equal to $5.48\times10^{-6}$ and $3.80\times10^{-5}$ for the $Z_\mathrm{L}$ and $Z_\mathrm{S}$ networks, respectively.

Each network is trained using the Levenberg-Marquardt algorithm described in \citet{demuth2014neural}.
The learning dynamics (i.e., evolution of the mean squared error for training, validation, and testing at different epochs of training procedure) are plotted in Figure~\ref{fig:reg_perf} for the largest and second largest remnants (left and right panels, respectively). At every training epoch, the weights of the networks are updated such that the mean squared error on the training data set gets progressively smaller. The mean squared errors converge in about 6 epochs for the surrogate model of the largest remnant and 4 for that of the second remnant.

Once trained, the predictive performance of the networks are quantified by the mean squared error (red curves in Figure \ref{fig:reg_perf}, Equation \ref{eq:MSE}) and correlation coefficient (box plots, Equation \ref{eq:R-value}) on the testing dataset. The mean squared error at convergence is equal to about $2\times10^{-4}$ with a correlation coefficient of above 96\% for the largest remnant, and $8\times10^{-4}$ with a correlation coefficient above 93\% for the second largest remnant.

\added{In Figure \ref{fig:reg_perf}, the differences between training, validation, and testing mean squared errors at convergence are smaller than 0.01 in units of $M_{core}/M_{planet}$. The mean squared errors on the testing set are smaller or equal to those on the training set, indicating that the trained algorithms generalize well the prediction to new cases not learned during training. The validation mean squared errors are also smaller than the training errors, which indicates that the trained algorithms are not overfitting the training datasets. These (small) differences between mean squared errors may also reflect differences in variance of the datasets, which we attempted to mitigate by populating the sets using random sampling without replacement.}

\added{Although the mean squared errors are globally low, inaccurate predictions may still occur locally in the parameter space. The distributions of the residuals in core mass fraction between $Q_{NN}$ and target values $Q_{SPH}$ for largest remnant and second remnant values, however, are well fit by Normal distributions $\mathcal{N} (0\%,2\%)$ and $\mathcal{N} (0\%,1\%)$, respectively. Despite the penury of data at large core mass fractions (i.e., $Z>40\%$), the surrogate model is found to be accurate in this regime too: the residuals are well fit with $\mathcal{N} (3\%, 4\%)$ and $\mathcal{N} (-3\%, 6\%)$ for the largest and second largest remnants, respectively. This noise level is comparable to the numerical uncertainty of the SPH simulations in predicting the core mass fraction of terrestrial planets, which we estimate to be 2--5 \% for accretionary and erosive events, and in the order of 10\% for erosive hit-and-run events. Predictions with residual $> 10\%$ account for just the 3.5\% and 7.8\% of the overall set for largest and second remnants respectively. As expected, we find that these inaccurate predictions tend to cluster in proximity of the boundary between hit-and-run and erosive collisions \citep[i.e., disruptive hit-and-runs,][]{2014NatGeoAsphaug}, in which substantial debris is produced in the erosion of both the targets' and projectiles' mantle, and where the SPH simulator is also expected to be the most noisy.} 

\subsection{Accretion efficiencies of the core and the mantle}
\label{sec:acc_effs_cm}

In order to track how the core mass fraction of a growing planet evolves through the giant impact phase of accretion, we split the change in mass from the target to the largest remnant into a core and mantle component:
\begin{equation}
\label{eq:change_L}
    M_\mathrm{L} - M_\mathrm{T} = \Delta M_\mathrm{L}^\mathrm{c} + \Delta M_\mathrm{L}^\mathrm{m},
\end{equation}
\noindent
where $\Delta M_\mathrm{L}^\mathrm{c} = M_\mathrm{L}^\mathrm{c} - M_\mathrm{T}^\mathrm{c}$ and $\Delta M_\mathrm{L}^\mathrm{m} = M_\mathrm{L}^\mathrm{m} - M_\mathrm{T}^\mathrm{m}$ are the changes in mass of the core and mantle of the largest remnant from the target as indicated by the superscripts ``c'' and ``m'', respectively.
Similarly, in the hit-and-run collision regime, we do the same for the change in mass from the projectile to the second largest remnant:
\begin{equation}
\label{eq:change_S}
    M_\mathrm{S} - M_\mathrm{P} = \Delta M_\mathrm{S}^\mathrm{c} + \Delta M_\mathrm{S}^\mathrm{m},
\end{equation}
where $\Delta M_\mathrm{S}^\mathrm{c} = M_\mathrm{S}^\mathrm{c} - M_\mathrm{P}^\mathrm{c}$ and $\Delta M_\mathrm{S}^\mathrm{m} = M_\mathrm{S}^\mathrm{m} - M_\mathrm{P}^\mathrm{m}$ are defined similarly as the largest remnant case.

The remnant bodies after a giant impact may have different core mass fractions than either of the pre-impact bodies (i.e. target and projectile) or each other.
For instance, a projectile may erode mantle material from the target in a hit-and-run collision, but during the same impact the target may accrete core material from the projectile.
In order to quantify and study these possibilities, we define distinct accretion efficiencies for each remnants' core ($\xi_\mathrm{L}^\mathrm{c}$ and $\xi_\mathrm{S}^\mathrm{c}$) and mantle ($\xi_\mathrm{L}^\mathrm{m}$ and $\xi_\mathrm{S}^\mathrm{m}$), so that, after dividing through by the projectile mass for normalization, the above expressions are transformed into:
\begin{align}
\label{eq:ML_segments} 
    \xi_\mathrm{L} &= Z_\mathrm{P} \xi_\mathrm{L}^\mathrm{c} + (1 - Z_\mathrm{P}) \xi_\mathrm{L}^\mathrm{m}, \\
\label{eq:MS_segments}     
    \xi_\mathrm{S} &= Z_p \xi_\mathrm{S}^\mathrm{c} + (1 - Z_\mathrm{P}) \xi_\mathrm{S}^\mathrm{m},
\end{align}
where we define the core and mantle component accretion efficiencies in the same manner as the overall accretion efficiencies of the remnant bodies:
\begin{align}
\label{eq:core_acc_L} 
    \xi_\mathrm{L}^\mathrm{c} &= \frac{M_\mathrm{L}^\mathrm{c}-M_\mathrm{T}^\mathrm{c}}{M_\mathrm{P}^\mathrm{c}} = \frac{Z_\mathrm{L}}{Z_\mathrm{P}}\xi_\mathrm{L} + \frac{Z_\mathrm{L}-Z_\mathrm{T}}{Z_\mathrm{P}}\frac{1}{\gamma}, \\
\label{eq:mantle_acc_L}
    \xi_\mathrm{L}^\mathrm{m} &= \frac{M_\mathrm{L}^\mathrm{m}-M_\mathrm{T}^\mathrm{m}}{M_\mathrm{P}^\mathrm{m}} = \frac{1-Z_\mathrm{L}}{1-Z_\mathrm{P}}\xi_\mathrm{L} - \frac{Z_\mathrm{L}-Z_\mathrm{T}}{1-Z_\mathrm{P}}\frac{1}{\gamma}, \\
\label{eq:core_acc_S}
    \xi_\mathrm{S}^\mathrm{c} &= \frac{M_\mathrm{S}^\mathrm{c}-M_\mathrm{P}^\mathrm{c}}{M_\mathrm{P}^\mathrm{c}} = \frac{Z_\mathrm{S}}{Z_\mathrm{P}}(1+\xi_\mathrm{S}) - 1, \\
\label{eq:mantle_acc_S}
    \xi_\mathrm{S}^\mathrm{m} &= \frac{M_\mathrm{S}^\mathrm{m}-M_\mathrm{P}^\mathrm{m}}{M_\mathrm{P}^\mathrm{m}} = \frac{1-Z_\mathrm{S}}{1-Z_\mathrm{P}}(1+\xi_\mathrm{S}) - 1,
\end{align}
where on the right-hand sides of Equations \ref{eq:core_acc_L}--\ref{eq:mantle_acc_S}, we express the core and mantle accretion efficiencies in terms found in Table~\ref{tab:data}: the initial projectile-to-target mass ratio ($\gamma$), the overall accretion efficiencies of the remnants ($\xi_\mathrm{L}$ and $\xi_\mathrm{S}$), and the core mass fractions of the final bodies ($Z_\mathrm{L}$ and $Z_\mathrm{S}$).

\added{In the SPH simulations discussed in Section \ref{sec:SPH_data}, the core mass fractions of the initial bodies are always $Z_\mathrm{T}=Z_\mathrm{P}=30\%$. The approach adopted for cases of collisions between bodies with core mass fraction different from $30\%$ --- which are expected to occur in planet formation studies --- is discussed in Section \ref{subsec:collision}.}

\section{Planetary differentiation model}
\label{sec:equi_method}

After a giant impact, the surviving mantle of a remnant body is assumed to equilibrate with any accreted material, in a magma ocean produced by the energetic impact.
This equilibration establishes the composition of the cooling magma ocean and also potentially involves dense Fe-rich metallic liquids that segregate to the core due to density differences.
In this Section, we describe how the inefficient-accretion model of Section \ref{sec:ML} is implemented into the planetary accretion and differentiation model published by \citet{2011Rubie,Rubie2015,2016Rubie}.
We direct the reader to the manuscripts by Rubie et al.\ for a more detailed description of the metal-silicate equilibration approach itself.

\subsection{Identification of silicate and metallic reservoirs}
\label{subsec:identification}

\added{Regardless of the impact conditions, in the perfect-merging model the metallic core of the projectile plunges into the target’s magma ocean turbulently entraining silicate liquid in a descending plume \citep{2011Deguen, 2003Rubie}. In the inefficient-accretion model, if the collision is a hit-and-run or erosive, the metallic core of the projectile does not plunge into the target's magma ocean, but some of the collision energy may be still delivered to the core-mantle boundaries of the colliding pair, potentially inducing some mixing of the metal and silicate reservoirs.}

\added{As today, however, there are no clear recipes for the quantification of the degree of mixing at the core-mantle boundary and re-equilibration of such mixture in case of hit-and-run and erosive collisions. Here, we explore two end-member scenarios: (1) no mixing between the reservoirs and, hence, no silicate-metal re-equilibration; and (2) the metal and silicate reservoirs are fully mixed and the mixture re-equilibrates. It is important to note that these two end-member scenarios are both unlikely, especially the second one, and that the reality of the degree of mixing and equilibration lies in-between these two end-members. In particular, previous studies do not give much support for the idea of full re-equilibration \citep[e.g.,][]{2016LPINakajima,Carter2015apj}, as this would require the delivery of a large input of energy at the core-mantle boundary. We therefore use these two end-members to bound the problem for the sake of studying the effect of switching on/off re-equilibration on planetary differentiation.} 

\begin{figure}
	\centering
	\includegraphics[width=0.85\linewidth]{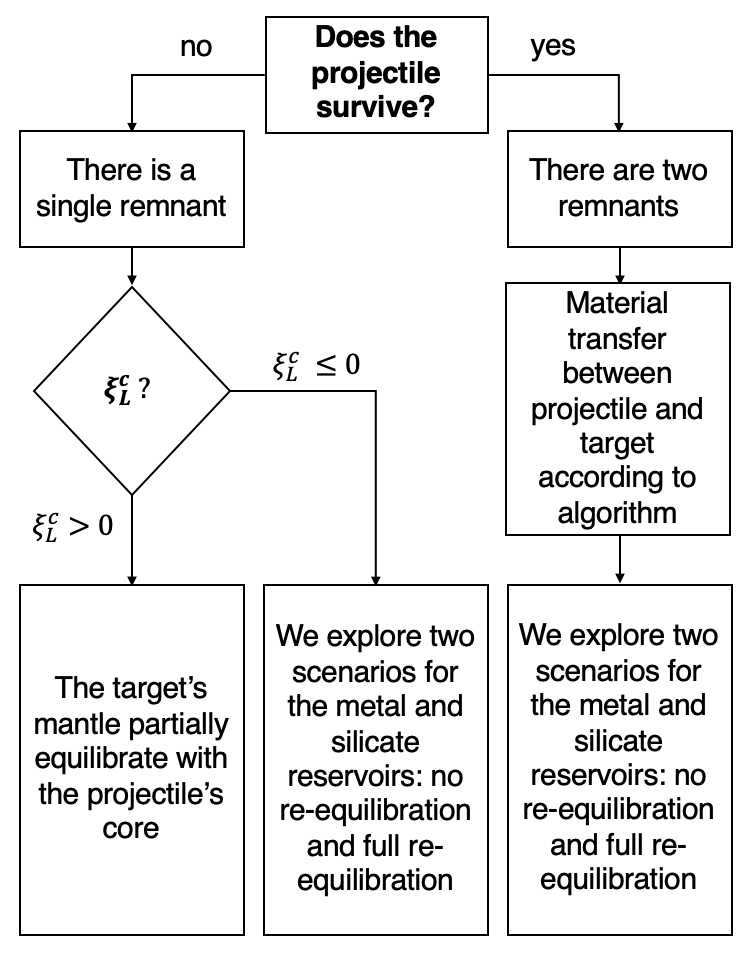}
	\caption{In the planetary differentiation model, we first identify the interacting reservoirs of the resulting bodies of a collision (the top three rows of this flowchart) using the surrogate model described in Section \ref{sec:ML}. The information is then passed to the metal-silicate equilibration model described in Section \ref{sec:equi_method} (bottom row).}
	\label{fig:flowchart}
\end{figure}

In the flowchart of Figure \ref{fig:flowchart}, we outline the steps for identifying equilibrating silicate and metal reservoirs from surviving target and accreted projectile material in the remnants' mantle. These steps are:
\begin{enumerate}
    \item The classifier of collision type (Section \ref{sec:class_reg}) is used to determine the number of resulting bodies of a collision.
    In case there is only a single remnant (accretion or erosion regimes), the core of the projectile may be either accreted or obliterated.
    This is determined by looking at the sign of the largest remnant's core accretion efficiency $\xi_\mathrm{L}^\mathrm{c}$.
    \item[2a.] If $\xi_\mathrm{L}^\mathrm{c}$ is positive and the event is not an hit-and-run collision, 
    the metallic core of the projectile plunges into the target's magma ocean turbulently entraining silicate liquid in a descending plume \citep{2011Deguen, 2003Rubie}.
    The plume's silicate content increases as the plume expands with increasing depth.
    This determines the volume fraction of metal $\phi_\mathrm{met}$ \citep{Rubie2015}: 
    \begin{equation}
    \label{eq:x_sil}
        \phi_\mathrm{met} =  \bigg(1+\frac{\alpha z}{r_0}\bigg)^{-3}
    \end{equation}
    \noindent
    where $\alpha = 0.25$ \citep{2011Deguen}, $r_0$ is the initial radius of the projectile’s core and $z$ is depth in the magma ocean. Equation~\ref{eq:x_sil} allows estimating the mass fraction of the embryo’s mantle material that is entrained as silicate liquid since the volume of the descending projectile core material is known.
    Following chemical equilibration during descent, any resulting metal is added to the proto-core and the equilibrated silicate is mixed with the fraction of the mantle that did not equilibrate to produce a compositionally homogeneous mantle \citep{Rubie2015}, under the assumption of vigorous mixing due to mantle convection. 
    \item[2b.] If $\xi_\mathrm{L}^\mathrm{c}$ is negative, the target's core and mantle are eroded and the projectile is obliterated. For this case, \added{we explore the two assumptions of no re-equilibration and full re-equilibration between the silicate and metal reservoirs of the largest remnant.}
    \item[3.] For hit-and-run collisions, the projectile survives accretion and becomes the second remnant of the collision.
    There may be mass transfer between the colliding bodies as predicted using the surrogate models of Section \ref{sec:regressor_accs} and Section \ref{sec:regressor_CMF}.\added{{For this case, we explore the two assumptions of no re-equilibration and full-equilibration between the silicate and metal reservoirs of the two resulting bodies.}}
\end{enumerate}

\subsection{Mass balance}
\label{subsec:mass_balance}
After each \added{accretionary giant impact and in case of full-equilibration for hit-and-run and erosive collisions}, there is re-equilibration between any interacting silicate-rich and metallic phases in the remaining bodies, which ultimately determines the compositions of the mantle and core.
The volume of interacting material is determined as described in Section~\ref{subsec:identification} and shown schematically in Figure~\ref{fig:flowchart}.
The planetary differentiation model tracks the partitioning of elements in the post-impact magma ocean between a silicate phase, which is modeled as a silicate-rich phase mainly composed of SiO$_2$, Al$_2$O$_3$, MgO, CaO, FeO, and NiO, and a metallic phase, which is modeled as a metal reservoir mainly composed of Fe, Si, Ni, and O.
Both the silicate-rich and metal phases include the minor and trace elements: Na, Co, Nb, Ta, V, Cr, Pt, Pd, Ru, Ir, W, Mo, S, C, and H.

The identified silicate-rich and metal phases equilibrate as described by the following mass balance equation:
\begin{equation}
\begin{split}
\label{Eq:thermo_system}
       \mathrm{Silicate~Liquid~}(1) + \mathrm{Metal~Liquid~}(1) \\ \Longrightarrow \mathrm{Silicate~Liquid~}(2) + \mathrm{Metal~Liquid~}(2)
\end{split}
\end{equation} 
where the flag ``1'' indicates the system before equilibration and the flag ``2'' indicates the system which is equilibrated under the new thermodynamic conditions of the resulting bodies of a collision.
Thus, all equilibrating atoms are conserved and the oxygen fugacity of the reaction is set implicitly. \added{The mass balance equations for the major elements (Si, Al, Mg, Ca, Fe, Ni, and O) are iteratively solved by combining them with experimentally-derived expressions for their metal-silicate partitioning behavior. We use the predictive parameterizations by \citet{2009GeCoAMann} for Si and most other siderophile elements, \citet{2008EPSLKegler} for Ni and Co and \citet{2010Frost} for O. For a detailed description of this coupled metal-silicate partitioning and mass balance approach, see \citet{2011Rubie}.}

In the chemical equilibrium of Equation \ref{Eq:thermo_system}, the partitioning of the elements into the core and mantle is controlled by the three parameters of the model: pressure $P_e$, temperature $T_e$, and oxygen fugacity $f_{O_2}$ of metal-silicate equilibration.
These are in turn a function of the type of collision and its accretion efficiencies (Section \ref{sec:acc_effs_cm}), which control how the reservoirs of the target and projectile interact. In the following, the treatment of these thermodynamic properties in the context of inefficient accretion and equilibration is described. \added{For the case of no re-equilibration, $P_e$,  $T_e$ and $f_{O_2}$ cannot be defined as there is not a chemical reaction or an equilibrium between co-existing metal and silicate of known compositions.}

\subsubsection{Equilibration pressure and temperature}
\label{sec:equi_TP}
Following each event \added{involving metal-silicate equilibration}, the silicate-rich and the iron-rich reservoirs are assumed to equilibrate at a pressure $P_\mathrm{e}$ which is a constant fraction $f_\mathrm{P}$ of the embryos' evolving core–mantle boundary pressure $P_\mathrm{CMB}$,
\begin{equation}
\label{eq:P_cmb}
    P_\mathrm{e} = f_\mathrm{P} P_\mathrm{CMB},
\end{equation} 
\noindent
and the equilibration temperature $T_\mathrm{e}$ is forced to lie midway between the peridotite liquidus and solidus at the equilibration pressure $P_\mathrm{e}$ \citep[e.g.,][]{Rubie2015}. 

The pressure $P_\mathrm{e}$  defined in Equation \ref{eq:P_cmb} is a simplified empirical parameter which averages the equilibration pressures for different types of impact events, and the constant $f_\mathrm{P}$ is a proxy for the average depth of impact-induced magma oceans. Here, we adopt a value $f_\mathrm{P} = 0.7$ for all the accretion events, consistently with the findings by \citet{2016PEPSdeVries}, which studied the pressure and temperature conditions of metal-silicate equilibration, after each impact, as Earth-like planets accrete. \added{In case of full-requilibration following hit-and-run and erosive collisions, we assume $f_\mathrm{P} = 1$ (that is,  $P_\mathrm{e} = P_\mathrm{CMB}$) because the metal and silicate reservoirs are in a mixed state and re-equilibrate at a pressure equal to that of the core-mantle boundary.}

For each of the resulting bodies, we determine the pressure $P_\mathrm{CMB}$ by using Equation 2.73 in \citet{2002geobook}.
The radial position of the core-mantle boundary is computed by using the approximation that an embryo is a simple two-layer sphere of radius $R$ consisting of a core of density $\rho_\mathrm{c}$ and radius $R_\mathrm{c}$ surrounded by a mantle of thickness ($R-R_\mathrm{c}$): 
\begin{equation}
\label{eq:core_radius}
    R_\mathrm{c} = \left(\frac{\rho -\rho_\mathrm{m}}{\rho_\mathrm{c}-\rho_\mathrm{m}}\right)^{\frac{1}{3}} R
\end{equation}
where the embryo's mean density $\rho$ is provided by the density-mass relationship introduced in \papertwo{} \added{normalized to predict the density of Earth (\SI{5510}{kg/m^3}) for a planetary mass of 1$M_\oplus$}.
Assuming a mantle density $\rho_\mathrm{m} = 0.4~\rho_\mathrm{c}$ (\footnote{The approximation that $\rho_\mathrm{m} = 0.4\rho_\mathrm{c}$ follows from the ratio between the densities of uncompressed peridotite and iron: $\rho_\mathrm{peridotite}/\rho_\mathrm{iron} = \SI{3100}{kg/m^3} / \SI{7874}{kg/m^3} \sim 0.4$.}), the core density $\rho_\mathrm{c}$ is equal to
\begin{equation}
\label{eq:dens_core}
    \rho_\mathrm{c} = \frac{5}{2} \rho \left(1-\frac{3}{5}Z\right)
\end{equation}
where $Z$ is the embryo's core mass fraction as predicted by the surrogate models in Section \ref{sec:regressor_CMF}.

\added{For an Earth-mass planet with core mass fraction of 30\%, the assumption of two constant density layers as presented above provides $P_\mathrm{CMB}$ = \SI{130}{GPa}, which is just 4.4\% lower than $P_\mathrm{CMB}$ of the modern Earth \citep[\SI{136}{GPa},][]{2002geobook}. In the immediate aftermath of giant impacts, the internal pressure may be lower than that of the modern value due to the heat release in the impact and higher rotation rates of the planet; the pressure is nevertheless expected to increase to its steady state value due to subsequent de-spinning, heat loss into space and vapor deposition \citep{2019SciLockStewart}.}

\subsubsection{Oxygen fugacity}
\label{sec:ox_fug}

The oxygen fugacity $f_\mathrm{O_2}$ determines the redox conditions for geologic chemical reactions. 
It is a measure of the effective availability of oxygen for redox reactions, and it dictates the oxidation states of cations like iron that have multiple possible valence states. For oxygen-poor compositions, oxygen fugacity is a strong function of \added{the equilibration} temperature because the concentration of Si in the metal strongly increases with temperature which increases the concentration of FeO in the silicate.
For more oxidized compositions, both Si and O dissolve in the metal, and oxygen fugacity is a much weaker function of temperature than in the case of more reduced bulk compositions.

The major benefit of the mass balance approach to modeling metal-silicate equilibration as described in Section \ref{subsec:mass_balance} is that the oxygen fugacity does not need to be assumed (as is done in most core formation models), but it is determined directly from the compositions of equilibrated metal and silicate (Equation \ref{Eq:thermo_system}).
The oxygen fugacity is defined as the partition coefficient of iron between metal and silicate computed relative to the iron-w\"{u}stite buffer (IW, the oxygen fugacity defined by the equilibrium 2Fe + O\textsubscript{2} = 2 FeO):
\begin{equation}
\label{eq:ox_fug}
\log{f_\mathrm{O_2}} (\Delta \mathrm{IW}) = 2 \log{\left( \frac{X_\mathrm{FeO}^\mathrm{Mw}}{X_\mathrm{Fe}^\mathrm{met}} \right)}
\end{equation}
\noindent
\added{where $X$ represents the mole fractions of components in metal or silicate liquids. $X_\mathrm{FeO}^\mathrm{Mw}$ is related to the silicate liquid composition \citep{2011Rubie} and $X_{Fe}^{met}$ is the fraction of total
iron in the bulk composition that is present as metal, as opposed to oxide (i.e., FeO in the
silicate). In Equation \ref{eq:ox_fug}, the activity coefficients are assumed to be unity because of the high temperatures, as is normally
done when calculating $f_{O_2}$ in studies of core formation, and because their values are very poorly known at high pressures and temperatures. This assumption is discussed in detail
by \citet{1999GeCoAGessmann}.}

\section{Inefficient accretion versus perfect merging: single impact events}
\label{sec:map_elements}

\begin{figure*}
	\centering
	\includegraphics[width=\linewidth]{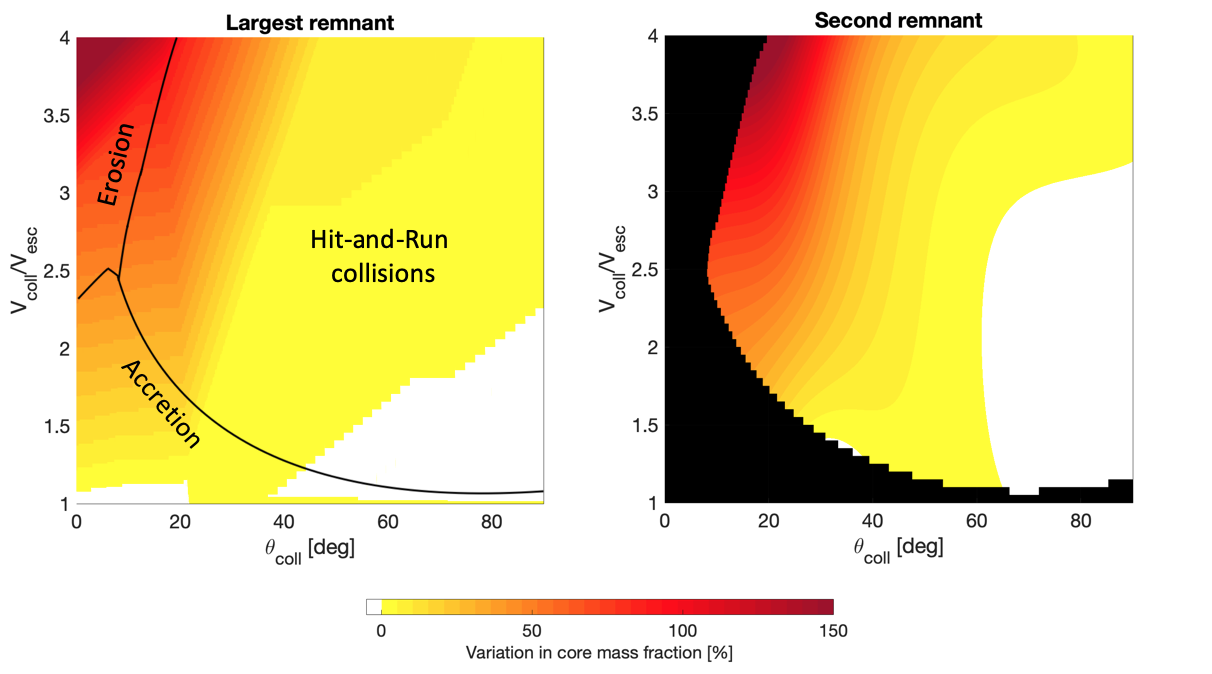}
	\caption{Relative difference between the core mass fractions of the largest remnant (left panel) and second remnant (right panel) and those of the parent bodies (the target's and projectile's, respectively) for a collision with $M_\mathrm{T}=\SI{0.1}{\mearth}$ and projectile-to-target mass ratio $\gamma = 0.7$ happening at different impact angles $\theta_{coll}$ and impact velocities $v_{coll}$ (in units of escape velocity $v_{esc}$). \added{The values are predicted using the surrogate model of core mass fraction of Section \ref{sec:regressor_CMF}.}
	Each point represents a different collision scenario in which the two bodies collide at different impact angles and velocities. The black curves define the different accretion regimes predicted using the classifier of Section \ref{sec:class_reg}. For comparison, the perfect-merging model predicts that the outcome of the collision is a single larger embryo with core mass fraction equal to that of the target (i.e.\ relative difference equal to zero) and no second remnant exists.}
	\label{fig:delta_CMF}
\end{figure*}

Here, we compare the predictions of the perfect-merging model and the inefficient-accretion model of Section \ref{sec:ML} for the case study of a single giant impact between a target of mass $M_\mathrm{T}=\SI{0.1}{\mearth}$ and a projectile of mass $M_\mathrm{P}=0.7 M_\mathrm{T}$. For the two models, we analyze the core mass fraction (Section \ref{subsec:single_CMF}), accretion efficiencies of the cores and the mantles (Section \ref{subsec:single_accs}), pressure, temperature, and oxygen fugacity of metal-silicate equilibration of the resulting bodies (Section \ref{sec:PTox}). \added{In the case of composition, we assume that the two embryos differentiated from early solar system materials that were chemically reduced. A high value of the fraction of total Fe in the system that is present in metal ($X_\mathrm{Fe}^\mathrm{met} = 0.99$) is justified in \citet{Rubie2015} as a condition to achieve reducing conditions so that elements such as Si and Cr partition sufficiently into the core. The refractory siderophile elements are assumed to be in solar-system relative proportions \citep[i.e. CI chondritic,][]{2003PalmeNeill}. This yields a core mass fraction which is approximately equal to 30\% for both target and projectile, which is that of the colliding bodies in the SPH data of Section \ref{sec:SPH_data}.}

\subsection{Core mass fraction}
\label{subsec:single_CMF}

\added{We show in Figure~\ref{fig:delta_CMF} that, in a collision between a target of mass $M_\mathrm{T}=\SI{0.1}{\mearth}$ and a projectile of mass $M_\mathrm{P}=0.7 M_\mathrm{T}$, the inefficient-accretion model predicts that the collision may result in two remnants and the resulting bodies' core mass fractions may be significantly different from those of their parent bodies depending on the characteristics of the collision (namely impact angle and velocity). In contrast, the perfect-merging model predicts that the outcome is a single, larger embryo of mass $M_\mathrm{T}+M_\mathrm{P}$ whose core mass fraction is equal to that of the target, i.e., its bulk composition is unchanged.}

\added{Cases of erosive collisions and disruptive hit-and-run events result in a net increase in the core mass fraction of the largest remnant due to massive amounts of mantle loss, which is predicted by our inefficient-accretion model trained on these events. In hit-and-run collisions, the projectile's core mass fraction is also larger than the pre-impact value, exacerbating the erroneous approximations of the perfect merging model, as the latter does not produce two diverse objects as a result of a single (hit and run) collision.}

\added{In two regions of the parameter space (white areas in Figure \ref{fig:delta_CMF}), the core mass fraction of the collision remnants is predicted to be at most 4\% and 3\% less than the pre-impact values. This corresponds to a variation of $\sim$ \SI{-1}{\%} in core mass fraction absolute value, which is within the noise floor of the surrogate models of core mass fraction (Figure \ref{fig:reg_perf}).}

\subsection{Accretion efficiency}
\label{subsec:single_accs}

\begin{figure*}
    \centering
	\includegraphics[width = 0.7\linewidth]{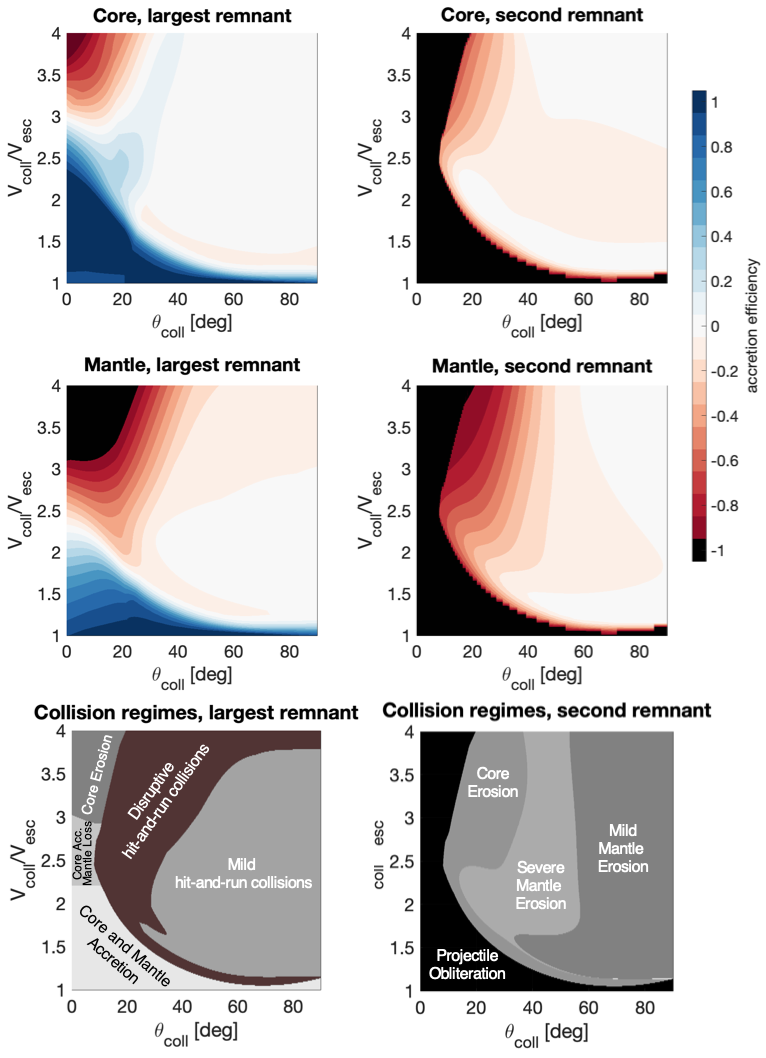}
	\caption{Accretion efficiency of the core (first row) and mantle (second row) for the largest remnant (left panels) and second largest remnant (right panels) and corresponding collision regimes as predicted by the inefficient-accretion model (third row) for a collision with $M_\mathrm{T}=\SI{0.1}{\mearth}$ and projectile-to-target mass ratio $\gamma = 0.7$, for different impact angles $\theta_{coll}$ and impact velocities $v_{coll}$ (in units of escape velocity $v_{esc}$).}
	\label{fig:acc_core_mantle_LR}
\end{figure*}

\added{For any collision, geochemical modellers that use the perfect-merging assumption tend to approximate that the projectile's mantle and core accrete into the target's mantle and core, and undergo equilibration separately} \citep{Rubie2015}. For a collision between a target of mass $M_\mathrm{T}=\SI{0.1}{\mearth}$ and a projectile of mass $M_\mathrm{P}=0.7 M_\mathrm{T}$,
the perfect-merging model predicts that the projectile's core plunges into the target mantle and that the entire projectile's mantle is accreted ($\xi_\mathrm{L}^\mathrm{m} = \xi_\mathrm{L}^\mathrm{c} = 1$), for every combination of impact angle and velocity.
\added{In contrast, our results show that there is a much larger diversity of core accretion efficiencies as a function of impact parameters, as shown in Figure~\ref{fig:acc_core_mantle_LR}, which demonstrates a critical inaccuracy of perfect merging and equilibration assumptions.}

In the parameter space of impact angle and impact velocity, we identify five collision regimes according to the core and mantle's accretion efficiencies of the largest remnant.
These are described in the bottom-left panel of Figure \ref{fig:acc_core_mantle_LR} and defined as:

\begin{enumerate}[label=\Alph*:]
    \item Core and mantle accretion occurs when $\xi_\mathrm{L}^\mathrm{c}\geq$ 0 and $\xi_\mathrm{L}^\mathrm{m}\geq$ 0. 
    \added{No second remnant is present because}
    the projectile plunges into the target and gets accreted.
    \item Core accretion with loss of mantle material occurs when $\xi_\mathrm{L}^\mathrm{c}\geq$ 0 and $\xi_\mathrm{L}^\mathrm{m}<$ 0.
    \added{No second remnant is present because} the projectile's core plunges into the target and gets accreted.
    \item Core erosion occurs when $\xi_\mathrm{L}^\mathrm{c}<0$.
    The target's mantle is catastrophically disrupted and core erosion may also occur.
    The largest remnant has a larger core mass fraction than the target (Figure \ref{fig:delta_CMF}).
    \added{No second remnant is present because} the projectile is obliterated.
    
    \item Mild hit-and-run collisions occur when $\xi_\mathrm{L}^\mathrm{c}\in[-0.1,~0.1]$ and $\xi_\mathrm{L}^\mathrm{m} \in[-0.1,~0.1]$.
    The target's core does not gain or lose substantial mass, while the target's mantle may lose some mass depending on the impact velocity.
    The bulk projectile escapes accretion and becomes the second remnant. Substantial debris production may occur.
    \item \added{Disruptive hit-and-run collisions occur in the rest of the parameter space.
    The target's loses part of its mantle and
    the largest remnant has a larger core mass fraction than the target (Figure \ref{fig:delta_CMF}).
    The bulk projectile escapes accretion and becomes the second remnant.}
\end{enumerate}

For the second remnant, we identify four collision regimes which are described in the bottom-right panel of Figure \ref{fig:acc_core_mantle_LR}:
\begin{enumerate}[label=\Alph*:]
    \setcounter{enumi}{5} 
    \item Mild mantle erosion occurs when $\xi_\mathrm{S}^\mathrm{c}\in[-0.1,~0.1]$ and $\xi_\mathrm{S}^\mathrm{m}\in[-0.1,~0.1]$.
    At high impact angle, the geometry of the impact prevents almost any exchange of mass between the target and the projectile.
    \item Severe mantle erosion occurs when $\xi_\mathrm{S}^\mathrm{c}\in[-0.1,~0.1]$ and $\xi_\mathrm{S}^\mathrm{m}<-0.1$.
    The second remnant has a less massive mantle compared to the projectile, while it retains its core mostly intact.
    \item Core erosion occurs when $\xi_\mathrm{S}^\mathrm{c}<$-0.1.
    The second remnant's core mass fraction is strongly enhanced with respect to that of the projectile.
    In disruptive hit-and-run collisions, the energy of the impact may be high enough to erode some core material.
    \item Projectile obliteration occurs when $\xi_\mathrm{S} = \xi_\mathrm{S}^\mathrm{c} = \xi_\mathrm{S}^\mathrm{m} = -1$.
    No second remnant exists, as the projectile is either accreted or completely disrupted.
\end{enumerate}
\smallskip

\added{In Figure \ref{fig:acc_core_mantle_LR}, we also observe that the surrogate model can produce nonphysical/inconsistent predictions for certain combinations of parameters. For example, at high angle and low velocity the surrogate model predicts that the core is eroded with accretion efficiency $\sim -0.1$ but mantle has an accretion efficiency near 0. This region is expected to be an artifact, since core erosion is expected to be accompanied by substantial mantle loss. This discrepancy, however, is within the noise floor of the surrogate model, which is estimated to be $\sim 0.15$ in units of accretion efficiency after error propagation.}

\subsection{Pressure, temperature and oxygen fugacity}
\label{sec:PTox}

\begin{figure}
\centering
	\includegraphics[width=\linewidth]{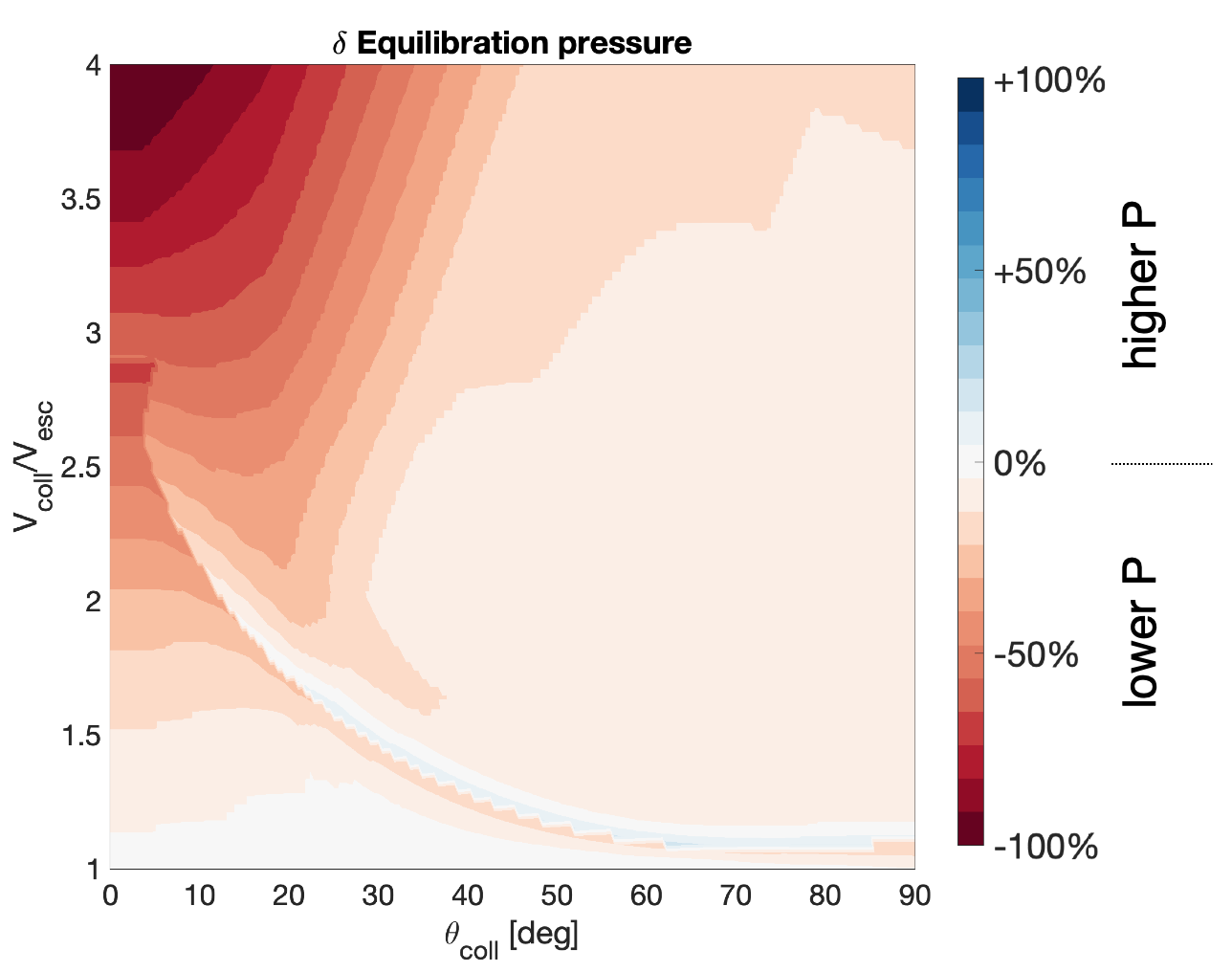}
	\includegraphics[width=\linewidth]{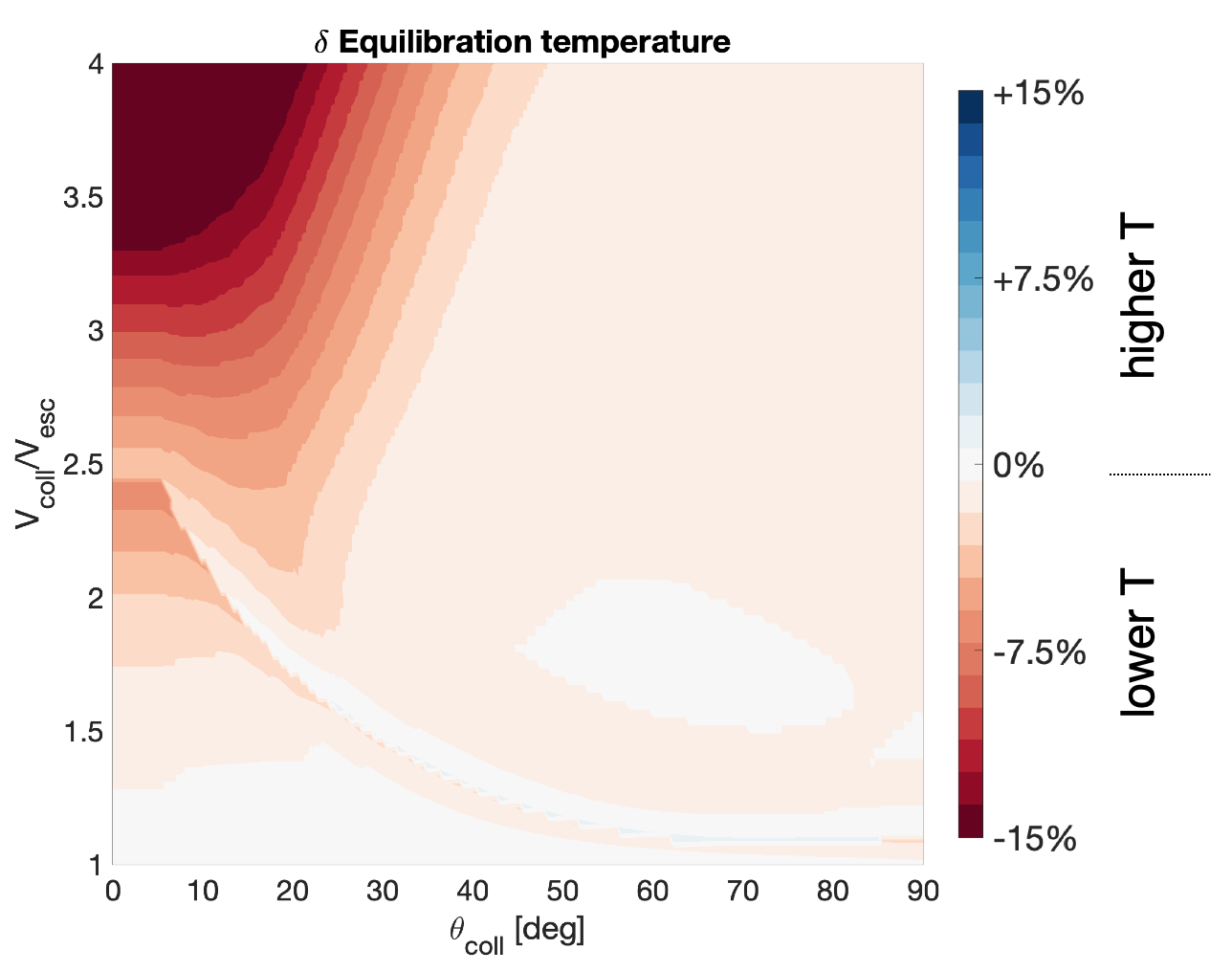}
	\includegraphics[width=\linewidth]{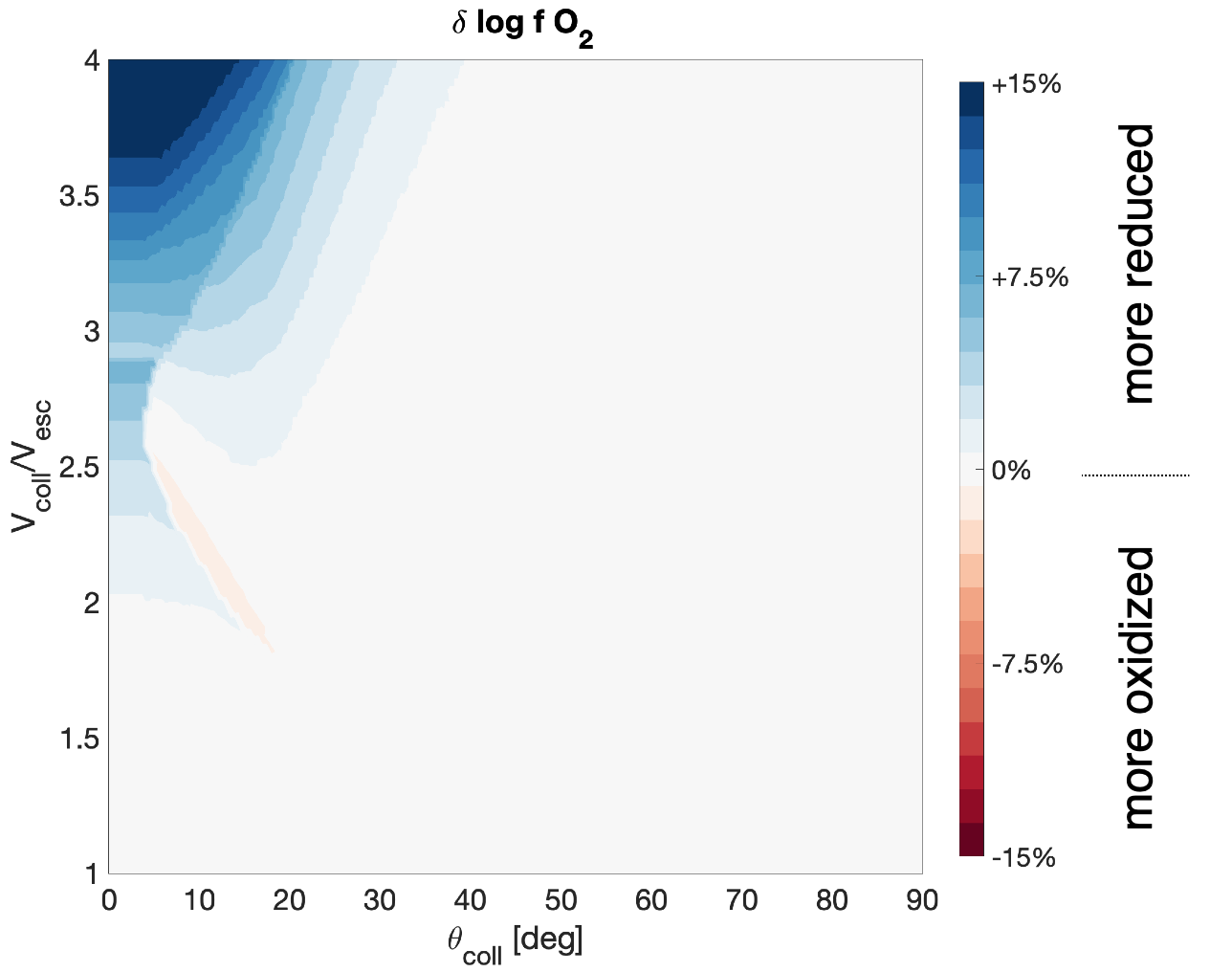}
	\caption{Relative difference (Equation \ref{eq:delta_X_PM}) between the predicted values of pressure, temperature, and oxygen fugacity of metal-silicate equilibration of the largest remnant from the inefficient-accretion model and those of the resulting body from the perfect-merging model (Equation \ref{eq:delta_X_PM}), for the same collision of Figure \ref{fig:delta_CMF} and \ref{fig:acc_core_mantle_LR}. The colliding embryos have initial molar fraction of iron in the metal phase $X_\mathrm{Fe}^\mathrm{met} = 0.99$. \added{We assume the full-equilibration scenario in case hit-and-run and erosive collisions (Section \ref{subsec:identification})}.}
	\label{fig:LR_vs_PM}
\end{figure}

The equilibration conditions of planets resulting from inefficient accretion (due to the stripping of mantle and core materials) will differ from those produced by the perfect-merging model. In Figure \ref{fig:LR_vs_PM} we show the difference in equilibration conditions for a single impact scenario (the same impact scenario as the previous sections); \added{hit-and-run and erosive collisions are treated assuming full re-equilibration, for which the equilibration conditions are well-defined}. The relative differences in the equilibration results are computed with respect to the perfect-merging values as
\begin{equation}
\label{eq:delta_X_PM}
    \delta X = \frac{X_\mathrm{inefficient ~accretion}-X_\mathrm{perfect~ merging}}{X_\mathrm{perfect~merging}} \times \SI{100}{\percent},
\end{equation}
\noindent
where $X$ indicates one of the thermodynamic metal-silicate equilibration parameters:  pressure, temperature, or oxygen fugacity of metal-silicate equilibration for an embryo with initial $X_\mathrm{Fe}^\mathrm{met} = 0.99$.
Positive values of $\delta[\log{f_\mathrm{O_2}}]$ indicate that the metal-silicate equilibration in the resulting body occurs at more chemically reduced conditions than in the planet from the perfect-merging model because of lower equilibration temperatures, while negative values of $\delta[\log{f_\mathrm{O_2}}]$ indicate  less chemically reduced conditions.

\added{As a result of mass loss in the target, the inefficient-accretion predictions deviate substantially from the perfect-merging predictions in case of erosive collisions (regimes B, C, D in in Section \ref{subsec:single_accs}) and disruptive hit-and-run collisions (regime E). Conversely, when the collision is accretionary (regime A), the equilibration conditions predicted by the inefficient-accretion model are similar to those obtained with the perfect-merging model ($\delta P_e \approx \delta T_e \sim \SI{0}{\percent}$ and $\delta[\log{f_\mathrm{O_2}}] \sim \SI{0}{\percent}$). Importantly, since the equilibration temperature is defined as a simple one-to-one function with respect to the equilibration pressure, as described in Section~\ref{sec:equi_TP}, the contours for temperature and pressure are very similar.}

\section{Inefficient accretion versus perfect merging: \textit{N}-body simulations}
\label{sec:summary_N_body}

In this Section, we investigate how planetary differentiation is affected by inefficient accretion during the end-stage of terrestrial planet formation. In contrast to Section \ref{sec:map_elements}, in which we investigate the case study of a single giant impact, here we use the core-mantle differentiation model to interpret the results of the \textit{N}-body simulations of accretion by \papertwo~which model the evolution of hundreds of Moon-to-Mars-sized embryos as they orbit the Sun and collide to form the terrestrial planets.
The goal of our analysis, however, is not to reproduce the solar system terrestrial planets, but to investigate whether or not the perfect-merging and inefficient-accretion models produce significantly different predictions for the terrestrial planets' final properties \added{after a series of collisions under a fiducial \textit{N}-body setup}.

\subsection{Initial mass and composition of the embryos}
\label{subsec:initial}

We use the data set of \textit{N}-body runs performed in \papertwo{} and test the effects of the two collision models under a single equilibration model. The dataset consists of 16 simulations that use the more realistic treatment of collisions (inefficient-accretion model) and, in addition, 16 control simulations where collisions are taken to be fully accretionary (perfect-merging model).
All the simulations were obtained with the \texttt{mercury6} \textit{N}-body code \citep{1999MNRASChambers}. For the inefficient-accretion model, \papertwo~use the code library  \texttt{collresolve}\footnote{\url{https://github.com/aemsenhuber/collresolve}} \citep{2019SoftwareEmsenhuberCambioni}. 
Each \textit{N}-body simulation begins with 153--158 planetary embryos moving in a disk with surface density similar to that for solids in a minimum mass solar nebula \citep{1977ApSSWeidenschilling}.
As in \citet{2001IcarusChambers}, two initial mass distributions are examined: approximately uniform masses, and a bimodal distribution with a few large (i.e., Mars-sized) and many small (i.e., Moon-sized) bodies.
The embryos in the simulations 01--04 all have the same initial mass \added{($\num{\sim1.67e-2} M_\oplus$)}.
In simulations 11--14, the embryos have their initial mass proportional to the local surface density of solids\added{; the minimum mass is $\num{1.59e-3} M_\oplus$}.
In simulations 21--24, the initial mass distribution is bimodal and the two populations of embryos are characterized by bodies with the same mass; \added{the minimum masses in the two populations are $\num{1.79e-3} M_\oplus$ and $\num{7.92e-2} M_\oplus$}.
In simulations 31--34, the initial mass distribution is also bimodal but the bodies have mass proportional to the local surface density\added{; the minimum embryo mass is $\num{2.14e-3} M_\oplus$}.

The compositions of the initial embryos are set by the initial oxygen fugacity conditions of the early solar system materials which are defined as a function of heliocentric distance.  Among the models of early solar system materials that have heritage in the literature, we adopt the model by \citet{Rubie2015}, whose parameters were refined through least squares minimization to obtain an Earth-like planet with mantle composition close to that of the Bulk Silicate Earth.
Accretion happens from two distinct reservoirs of planet-forming materials: one of reduced material in the inner solar system \SI{<0.95}{\au}, with the fraction of iron dissolved in metal equal to \num{0.99} and fraction of available silicon dissolved in the metal equal to \num{0.20}. Exterior to that, no dissolved silicon is present in the metal and the proportion of Fe in metal decreases linearly until a value of \num{0.11} is reached at \SI{2.82}{\au}. Beyond 2.82 au, the iron metal fraction linearly decreases and reaches about zero at \SI{6.8}{\au} \added{ \citep[see Figure 6 in][]{Rubie2015}.} 

\added{Within each group of \textit{N}-body simulations from \papertwo, which vary different aspects of the initial conditions, we compare results from the perfect merging and inefficient accretion scenarios.}  In sets 01--04 and 21--24, the initial mass \added{of every embryo is identical}, so the number density of embryos scales with the surface density. In sets 11--14 and 31--34, the initial embryos' masses scale with the local surface density; the spacing between embryos is independent of the surface density, but the heliocentric distance between them gets smaller as distance increases, hence the number density of the embryos increases with heliocentric distance. This means that the simulations 11--14 and 31--34 are initialized with most of the embryos forming farther from the Sun than those in simulations 01--04 and 21--24. Following the model by \citet{Rubie2015}, this implies that most of the embryos in simulations 11--14 and 31--34 form with initial core mass fractions \added{smaller than 30\%}.

\subsection{Working assumptions}
\label{subsec:collision}

\begin{enumerate}
    
    \item The \textit{N}-body simulations of \papertwo{} are based on the assumption that the bodies are not spinning prior to each collision and are not spinning afterwards.
    We acknowledge that this approximation violates the conservation of angular momentum \added{in off-axis collisions} and that collisions between spinning bodies would alter accretion behavior \citep[e.g.,][]{1999Agnor}.

    \item In the \textit{N}-body simulations with inefficient accretion by \papertwo, only the remnants whose mass is larger than \SI{1e-3}{\mearth} are considered. \added{Removing small bodies from the \textit{N}-body simulations avoids uncertainties that may arise from querying predictions from the machine learning model in regimes on which is was not trained (the SPH dataset extends down to collisions with a total mass of \SI{2e-3}{\mearth}). If the surrogate model predicts a mass smaller than the threshold, then this body is unconditionally treated as debris and does not dynamically interact with the embryos. \papertwo{} nevertheless tracked the evolution of the overall debris mass budget.}

    \item  As the embryos evolve during accretion through collisions, their core mass fractions can evolve to be different from 30\%, which is the core mass fraction of the SPH colliding bodies that were used to train the surrogate models in Section~\ref{sec:regressor_CMF}.
    For this reason, in the analysis of the \textit{N}-body simulations we make the approximation that the core mass fraction of each collision remnant \added{\textit{prior} to metal-silicate equilibration} is equal to
    \begin{equation}
    \label{eq:core_approx}
        Z \approx (Z^* - 30\%) + Z_0
    \end{equation}
    \noindent
    where $Z^*$ is the core mass fraction of the remnant as predicted by the surrogate model of Section~\ref{sec:regressor_CMF}, and $Z_0$ is the metal fraction of the parent body as computed by the core-mantle differentiation model. \added{To guarantee the conservation of siderophile material, the prediction by Equation \ref{eq:core_approx} is bounded to lie between 0 and 100\%.}
\end{enumerate}

\subsection{Results: Core mass fraction}
\label{subsec:N_body_res}

To compare the effect of the two accretion model assumptions (inefficient accretion and perfect merging) for planets in the \textit{N}-body simulations, we examine the final core mass fractions.
\added{We chose not to bin the data as we found the results to be highly sensitive to choices in bin width (low-\textit{N} statistical issues).}

\begin{figure*}
	\centering
	\includegraphics[width=0.8\linewidth]{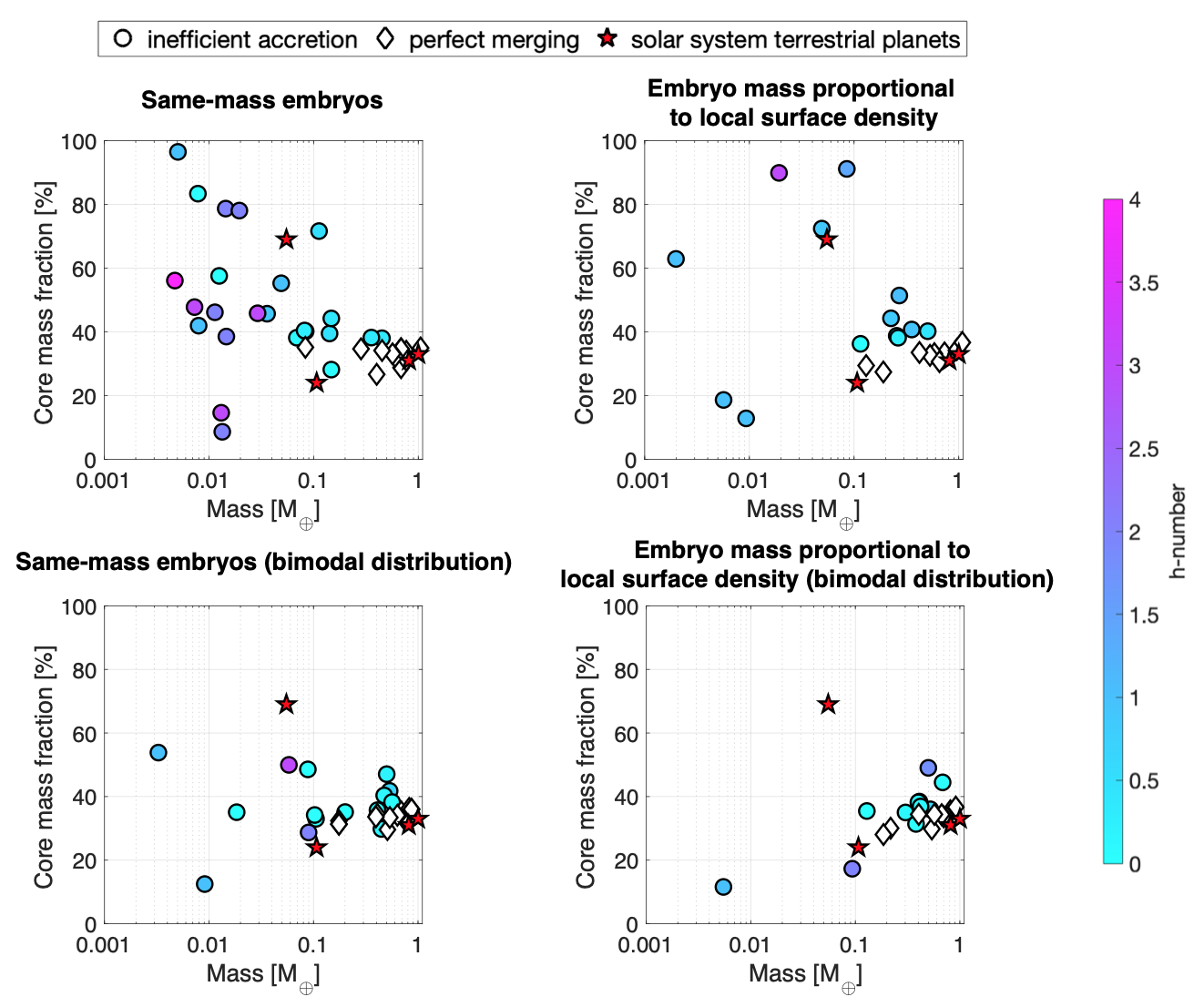}
	\caption{\textit{N}-body results by the inefficient-accretion model (dots) and the perfect-merging model (diamonds) in terms of the final planets' core mass fraction as function of the final planetary mass and different initial mass distributions for the embryos. \added{The core mass fractions of the inefficient-accretion planets are the arithmetic mean of the values obtained using the two assumptions of no re-equilibration and full re-equilibration of the metal and silicate reservoirs after hit-and-run and erosive collisions (Section \ref{subsec:identification})}. The red stars show the core mass fractions of the solar system terrestrial planets.}
	\label{fig:delta_ZZ}
\end{figure*}

\begin{figure*}
	\centering
	\includegraphics[width=\linewidth]{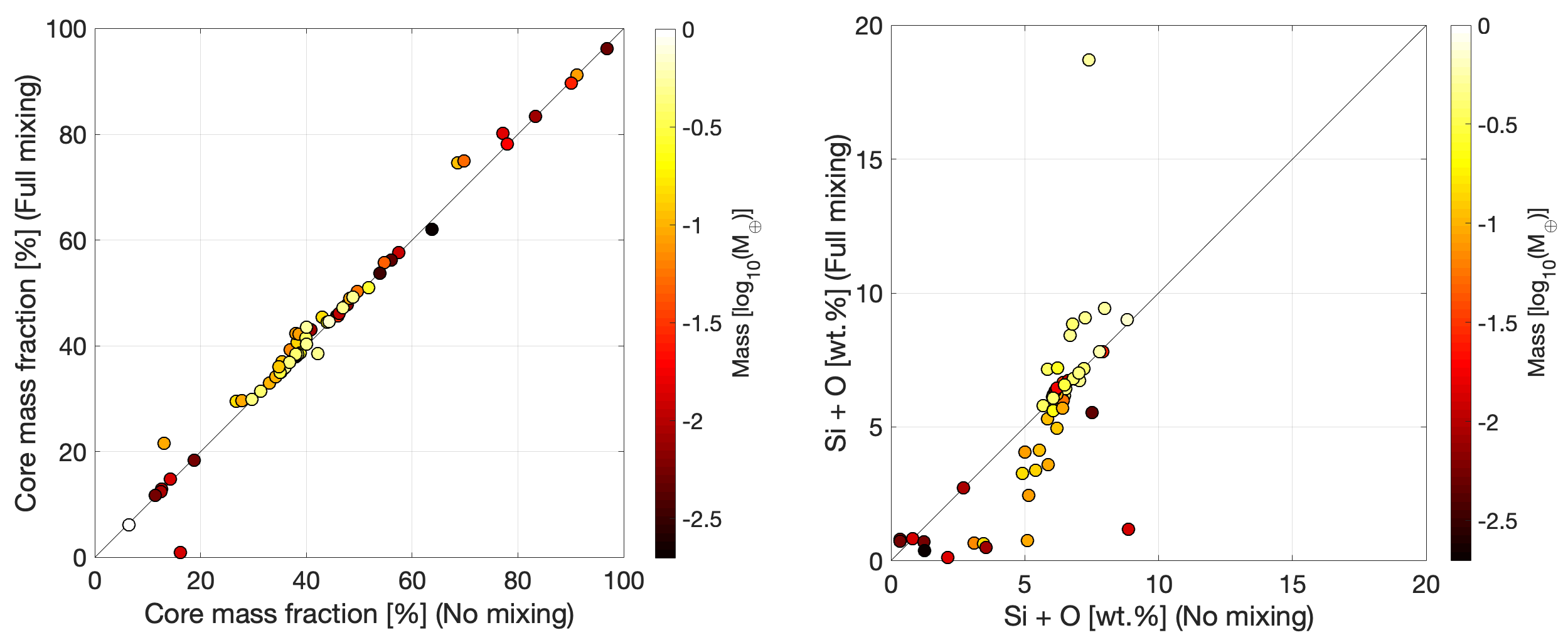}
	\caption{Left panel: core mass fraction of the terrestrial planets obtained in all the \textit{N}-body simulations from \papertwo~ assuming that full mixing and re-equilibration occur at the core-mantle boundary (vertical axis) or no re-equilibration occurs (horizontal axis) in hit-and-run and erosive collisions. Right panel: same as the left panel, but for the concentration of Si and O in the core of the terrestrial planets. The data points are color coded in terms of the planets' masses (logarithmic scale).}
	\label{fig:correlation}
\end{figure*}

Figure~\ref{fig:delta_ZZ} is a plot of the core mass fraction of the final planets as a function of planetary mass as determined by the perfect-merging model (open diamonds) and the inefficient-accretion model (dots).
Each of the four subpanels of Figure~\ref{fig:delta_ZZ} plots one of the four groups of \textit{N}-body simulations (01--04, 11--14, 21--24, 31--34) described in Section \ref{subsec:initial}.
Each subpanel shows the results from all the four \textit{N}-body simulations of the correspondent group.
The inefficient-accretion results are color-coded in terms of their $h$-number, which measures how many hit-and-run collisions an embryo experienced during the accretion simulation \added{(\citealp{2014NatGeoAsphaug}; \papertwo)}.
If the collision event is a merger, the largest remnant's $h$-number is equal to the mass average of the target's and projectile's $h$-numbers.
If the event is a hit-and-run collision, the second remnant's $h$-number is increased by 1, \added{while the $h$-number of the largest remnant does not change}.
We also plot the estimated values for the core mass fractions of the inner solar system terrestrial bodies as red stars: $Z_\mathrm{Earth} = \SI{33}{\percent}$ , $Z_\mathrm{Venus} = \SI{31}{\percent}$, $Z_\mathrm{Mars} = \SI{24}{\percent}$, $Z_\mathrm{Mercury} = \SI{69}{\percent}$ \citep[respectively]{2013EarthCMF,RUBIE201543,2017HelffrichMarsCore,2013Hauck}.

\added{Perfect merging simulations generally produce planets with core mass fractions close to 30\% regardless of the initial embryo distribution. The most massive planets produced by inefficient accretion also have core mass fractions near 30\%, but the degree of spread in core mass fraction increases substantially among the less massive bodies. Furthermore,} remnants with core mass fractions above 40\% are generally found to have relatively high $h$-numbers, meaning that they survived multiple hit-and-run collisions.

As discussed in \papertwo, the initial mass of the embryos influences the dynamical environment and thus imparts a change in the degree of mixing between feeding zones in the planetary disk.
This is found to affect the spread in core mass fraction of the smaller embryos. The simulations that produced the results in the top panels of Figure \ref{fig:delta_ZZ} were initialized with embryos of similar mass.
As a result, these simulations are characterized by a predominance of collisions between similar-mass bodies which tend to result in more hit-and-run collisions \citep{2010ChEGAsphaug,2020ApJGabriel}. \added{By analyzing the evolution of the debris mass using Equations \ref{eq:core_acc_L}--\ref{eq:mantle_acc_S}, we find that the debris is composed mainly of the mantle material of the embryos, with the core material contributing up to 15\% when collisions are predominantly hit-and-runs in nature. This suggests that} many of the less massive planets are the ``runners'' from hit-and-run collisions (blue and magenta colored points), which managed to escape accretion onto the larger bodies but lost mantle material in the process, an effect predicted in \citet{2014NatGeoAsphaug}. This also explains why remnants with high $h$-numbers tend to also have higher core mass fractions. \added{Conversely, in \textit{N}-body simulations that are initialized with embryos of dissimilar mass (bottom panels of Figure \ref{fig:delta_ZZ}), the most massive bodies establish themselves dynamically early on, accreting smaller bodies in non-disruptive collisions. This leads to final bodies with relatively similar core mass fractions.} 

\added{Small planets that have a low \textit{h}-number (cyan colored dots) are either giant-impact-free embryos (i.e., small planets with core mass fraction $\sim 30\%$) or the largest remnants of disruptive collisions (i.e., small planets with core mass fraction $> 30\%$). In reality, the latter would sweep up some of their debris (an effect not included in the \textit{N}-body simulations by \papertwo); this would lower the value of their core mass fraction, as mantle material is preferentially eroded in collisions.} We also observe that the spread in core mass fraction in small bodies depends also on the heliocentric distance at which they form.
Adopting the model for early solar system materials by \citet{Rubie2015}, the embryos that form farther than \SI{0.95}{au} have an initial core mass fraction lower than 30\% due to oxidation of metallic iron by water.
This explains the presence of a few small embryos with core mass fraction lower than 30\%.  

\added{Finally, we find that the distribution of core mass fraction as function of mass is robust against the assumption of the degree of mixing and re-equilibration of the metal and silicate reservoirs in hit-and-run and erosive collisions. This is shown in Figure \ref{fig:correlation}, left panel, which is a plot of the the core mass fraction values of the inefficient-accretion planets predicted using the assumption of full mixing and re-equilibration (vertical axis) versus the results assuming no re-equilibration (horizontal axis). The predictions are well-fit by the 1:1 line with coefficient of determination $R^2$ = 0.98. This means that the diversity in core mass fraction observed in Figure \ref{fig:delta_ZZ} is primarily set by the erosive effect of giant impacts, and that subsequent re-equilibration of mixed reservoirs (or lack thereof) do not affect the prediction of core mass fraction. In contrast, the chemical composition of the metal and silicate reservoirs still depend upon the adopted assumption. This is shown in Figure \ref{fig:correlation}, right panel, which is a plot of the concentration of Si and O in the core of the inefficient-accretion planets predicted using the assumption of full mixing and re-equilibration (vertical axis) versus the results assuming no re-equilibration (horizontal axis). As opposed to the core mass fractions (left panel), the data points in the right panel of Figure \ref{fig:correlation} are not well-fit by the 1:1 line. This warrants future studies on the nature and extent of mixing and re-equilibration at the core-mantle boundaries, as further discussed in Section \ref{subsec:fut_work}}.

\added{\section{Discussion}}
\subsection{Statistics of planetary diversity}
 \begin{table*}
    \centering
        \caption{Average value and range of the values of core mass fractions for the two populations of Inefficient-Accretion (IA) planets with M$<$0.1 M$_\oplus$ and M$>$0.1 M$_\oplus$. The symbols $\bar{Z}$ and $\Delta Z$ indicate the average core mass fraction of the population and its range, respectively. For reference, the Perfect-Merging (PM) planets have $\bar{Z}_{PM}~\approx$ \SI{33}{\%} and $\Delta Z_{PM}~\approx$ \SI{8}{\%}.}
        \label{tab:comparison_01}
    \begin{tabular}{ccccc}
        \hline
         Sims & $\bar{Z}_{IA}$ (M $<$ 0.1 M$_\oplus$) & $\Delta Z_{IA}$ (M $<$ 0.1 M$_\oplus$) & $\bar{Z}_{IA}$ (M $>$ 0.1 M$_\oplus$) & $\Delta Z_{IA}$ (M $>$ 0.1 M$_\oplus$) \\
        \hline
        01--04 & 51\% & 88\%  &   40\%  &   16\%\\
        11--14 & 57\%  &  78\%  &  41\%  &  16\%\\ 
        21--24 & 38\%  &   41\%  &   37\%  &     17\%  \\
        31--34 & 15\%  &    6\%  &    38\%  &    18\% \\
        \hline
        \\
    \end{tabular}
\end{table*}

\added{In order to quantify the observed diversity in smaller bodies, we compute the mean core mass fraction and the range in core mass fraction (i.e., maximum minus minimum values) of the two sub-populations of inefficient-accretion planets that have mass larger and smaller than $0.1 M_\oplus$ (``more massive'' and ``less massive'' planets, respectively, Table~\ref{tab:comparison_01}), and compare them to those of the perfect-merging planets. We use a cut-off value of $0.1 M_\oplus$ because the perfect-merging simulations lack final planets smaller than $0.1 M_\oplus$ (with the exception of 1 planet in the simulations 01--04), and because $0.1 M_\oplus$ roughly corresponds to the maximum mass of the initial embryos of the \textit{N}-body simulations from \papertwo{}.}

\added{The statistics in Table \ref{tab:comparison_01} shows that the more massive inefficient-accretion planets tend to be fairly similar to one another in terms of core mass fraction, while there is a higher degree of diversity among the less massive planets. Futhermore, the more massive planets have core mass fraction statistics similar to that of the perfect-merging planets; this similarity is expected to be further enhanced if debris re-accretion is modelled, because debris tends to be preferentially accreted by the larger planets. Conversely, the average core mass fraction and its spread for the less massive planets tend to be higher than those of the perfect-merging planets because the former are preferentially eroded by giant impacts \citep{2010ChEGAsphaug}. This is especially true when collisions are predominantly hit-and-runs in nature (e.g., results in the top panels of Figure \ref{fig:delta_ZZ}).}

\subsection{Effect of debris re-accretion}
\label{sec:debris}

\added{In the \textit{N}-body simulations of \papertwo{}, debris produced in the aftermath of a giant impact that was not bound to the target or the runner was simply removed from the system. The mass of debris produced in a given giant impact is usually a relatively minor fraction of the colliding mass, compared to the major bodies, but cumulatively this simplification leads to significant mass removal from the system: up to 80\% of the mass in the initial embryos (\papertwo{}). Here we consider how this may affect our predictions of compositional diversity with size.}

\added{Other studies of terrestrial planet formation have treated giant impact debris explicitly} \citep[e.g.,][]{2009ApJStewartLeinhardt,2010ApJKokubo,2013IcarusChambers,2016ApJQuintana}\added{, although there is substantial difference from study to study in the treatment of debris production, the manner of its release into orbit, and the manner in which debris interacts gravitationally with itself and other bodies. In addition, there are differences in the calculation of the accretion efficiency of giant impacts.
\citet{2013IcarusChambers}, whose initial conditions were replicated in \papertwo{}, remains the most useful point of comparison, as it provides end-member cases of perfect merging, as well as simulations where debris is produced according to a scaling law \citep{2012ApJLeinhardt}. The orbital release of debris in the simulations from \citet{2013IcarusChambers} is represented as a few particles radiating isotropically from the collision point at 1.05 times the target escape velocity, although it is unclear whether the debris fragments subsequently interact with each other or only with the major bodies.}

\added{With these approximations, \citet{2013IcarusChambers} found that planets took longer to reach their final masses, due to the sweep up of simulated debris particles, than in the cases with perfect merging. The mean time required for Earth analogues to reach their final mass is \SI{\sim159}{Myr}, substantially longer than \SI{101}{Myr} under perfect merging. The simulations in \papertwo{} neglect debris but include realistic models of inefficient accretion, which results in a protracted tail of final accretion events lasting to \SI{\sim200}{Myr}. The longer timescale in this case is due to the realistic inclusion of hit-and-run collisions and the more accurate (stronger) orbital deflections for the runner in those simulations. Hit-and-run impactors in \papertwo{} appear to have served a similar role as the debris particles in \citet{2013IcarusChambers}.}

\added{Of greater importance for the present study are the masses and compositions of the finished planets. Compared to perfect merging, \citet{2013IcarusChambers} found that more terrestrial planets (4-5) are produced, with a wider distribution of resulting masses, when collisional fragmentation is approximated as described. Simulations in \papertwo{} had a median number of 3 and 4 final planets under the perfect merging and realistic-accretion assumptions, respectively. The standard deviation in the final number of planets in \papertwo{} is 0.7 for the perfect-merging simulations, and 2 when applying the inefficient-accretion model.}

\added{The evolution and influence of debris can also have important dynamical effects on the accreting planets. Similar to the effects noted in \citet{2006IcarusOBrien}, the simulated debris particles in \citet{2013IcarusChambers} appear to have applied a dynamical friction to the orbits of the major bodies, leading to slightly lower eccentricities than those of the perfect-merging simulations. In light of this, neglecting debris in \papertwo{} is likely to have allowed a greater dynamical excitation of the growing planets, which subsequently would have collided at higher impact velocities. Half of the collisions in \papertwo{} had $V_{coll} >$ 1.6 $V_{esc}$, and a quarter had $V_{coll} > $ 2.12 $V_{esc}$, whereas 95\% of the collisions in \citet{2013IcarusChambers} had velocities lower than 1.6$v_{\rm{esc}}$\footnote{The impact velocity distribution is provided in Supplementary Material of \cite{2020ApJGabriel}}. Further comparison of dynamical excitation and planet formation is difficult, because \citet{2013IcarusChambers} applied a hard cut-off to discern between accretion/disruption with debris production and hit-and-run collisions. In this simplified approach, regardless of impact velocity, a giant impact at impact angle greater than \ang{30} is treated as a hit and run, and there is no gravitational deflection of the runner. In \papertwo{}, by contrast, the close approaches and resulting angular momentum distribution between the colliding bodies are explicitly resolved. At relatively high velocity, most of the collisions at low impact angle predicted to be accretionary by \citet{2013IcarusChambers} are actually hit-and-run, as discussed in \paperone{}; conversely, at low velocity, many of the collisions predicted by \citet{2013IcarusChambers} to be hit-and-run are actually accretions.}

\added{Of paramount interest to the present work is how the approach of ignoring the debris influences our predictions for planetary diversity. Debris that are produced in a giant impact are on heliocentric orbits that can intersect the target, and can intersect the runner in the case of hit-and-run collisions. As argued by \citet{2014NatGeoAsphaug}, accretion occurs preferentially onto the most massive body, the post-impact target, which has the substantially larger collisional cross-section and sweeps up most of the debris. Re-impacts with the runner are less frequent, and furthermore are more likely to be erosive. Therefore, the inclusion of a proper treatment of debris-embryo interaction is expected to strengthen our conclusion, that planetary diversity increases at smaller scales. }

\subsection{Future work}
\label{subsec:fut_work}

\added{In future studies, we will develop a recipe for the degree of re-equilibration, in-between the two end members studied here, as a function of impact conditions for giant impacts, building upon what it has been already achieved in previous studies \citep[e.g.,][]{2010EPSLDahlStevenson,2016LPINakajima,raskin2019material}. This includes: (1) studying at which depth in the planet equilibration occurs, i.e., determining the value of $f_\mathrm{P}$ (Equation \ref{eq:P_cmb}) for different collision regimes; and (2) a more-advanced prediction for the equilibration temperature. In the planetary differentiation model, the latter is set at the midpoint between the solidus and liquidus temperatures of mantle peridotite. 
In this way, we have enforced that the temperature behaves like the pressure of metal-silicate equilibration. 
This assumption is probably accurate for the case of a projectile's core sinking through a mantle magma ocean, but proving its validity for hit-and-run or erosive collisions requires further study.}

\added{Future work will also aim to further improve the realism of the surrogate model of giant impacts. 
Specifically, we plan to train neural networks on SPH simulations of collisions between embryos with core mass fraction different from 30\% and refine the treatment of the debris field in \textit{N}-body simulations with respect to that used in \papertwo{}. 
As discussed in Section \ref{sec:debris}, in addition to increase the mass of the surviving embryos, debris are expected to reduce the eccentricities and inclinations of the embryos via dynamical friction \citep{2006IcarusRaymond,2006IcarusOBrien,2019ApJKobayashi}, thus increasing the accretion efficiency but reducing the interactions across feeding zones.}

\section{Conclusions}
\label{sec:conclusion}

In this work, neural networks trained on giant impact simulations have been implemented in a core-mantle differentiation model coupled with \textit{N}-body orbital dynamical evolution simulations to study the effect of inefficient accretion on planetary differentiation and evolution.
We make a comparison between the results of the neural-network model (``inefficient accretion'') and those obtained by treating all collisions unconditionally as mergers with no production of debris (``perfect merging''). 

For a single collision scenario between two planetary embryos, we find that the assumption of perfect merging overestimates the resulting bodies' mass and thus their equilibration pressure and temperature.
Assuming that the colliding bodies have oxygen-poor bulk compositions, the inefficient-accretion model produces a wider range of oxidation states that depends intimately on the impact velocity and angle; mass loss due to inefficient accretion leads to more reduced oxygen fugacities of metal-silicate equilibration because of the strong temperature dependence at low oxygen fugacities.

To investigate the cumulative effect of giant impacts on planetary differentiation, we use a core-mantle differentiation model to post-process the results of \textit{N}-body simulations obtained in \papertwo, where terrestrial planet formation was modeled with both perfect merging and inefficient accretion. \added{The inefficient-accretion model suggests that planets less massive than \SI{0.1}{\mearth} are compositionally diverse in terms of core mass fraction. This is driven by the effect of mantle erosion that is included in the machine-learning model}.
In contrast, both models provide similar predictions for planets more massive than \SI{\approx0.1}{\mearth}, \added{e.g., a tight clustering near 30--40\% value of core mass fraction.}
This is consistent with previous studies that successfully reproduced Earth's Bulk Silicate composition using the results from \textit{N}-body simulations with perfect merging \citep{Rubie2015,2016Rubie}.
We therefore suggest that an inefficient-accretion model is necessary to accurately track compositional evolution in terrestrial planet formation\added{, particularly when it comes to modeling the history and composition of less massive bodies. Our results improve upon previous studies that reached similar conclusions but did not find planets with core mass fraction comparable to that of Mercury \citep[e.g.,][]{2015IcarDwyer} or whose simulations (specifically the simplification of close approaches) likely underestimated mantle erosion \citep[e.g.,][]{Carter2015apj}.}

Finally, the value of oxygen fugacity of metal-silicate equilibration is known to influence the post-accretion evolution of rocky planets' atmospheres \citep[e.g.,][]{2007SSRZahnle,2019SciArmstrong,2020arXivZahnle}. Improving the realism of planet formation models with realistic collision models therefore becomes crucial not only for understanding how terrestrial embryos accrete, but also to make testable predictions of how some of them may evolve from magma-ocean planets to potentially habitable worlds. 

\acknowledgments
S.C., A.E., E.A., and S.R.S. acknowledge support from NASA under grant 80NSSC19K0817. We thank the anonymous reviewers for the comments and and edits that improved this manuscript. The authors also thank A. Morbidelli and M. Nakajima for discussions on the nature of terrestrial planet formation and metal-silicate equilibration.


\begin{thebibliography}{}
\expandafter\ifx\csname natexlab\endcsname\relax\def\natexlab#1{#1}\fi
\providecommand{\url}[1]{\href{#1}{#1}}
\providecommand{\dodoi}[1]{doi:~\href{http://doi.org/#1}{\nolinkurl{#1}}}
\providecommand{\doeprint}[1]{\href{http://ascl.net/#1}{\nolinkurl{http://ascl.net/#1}}}
\providecommand{\doarXiv}[1]{\href{https://arxiv.org/abs/#1}{\nolinkurl{https://arxiv.org/abs/#1}}}

\bibitem[{{Agnor} \& {Asphaug}(2004)}]{2004ApJAgnor}
{Agnor}, C., \& {Asphaug}, E. 2004, \apjl, 613, L157, \dodoi{10.1086/425158}

\bibitem[{{Agnor} {et~al.}(1999){Agnor}, {Canup}, \& {Levison}}]{1999Agnor}
{Agnor}, C.~B., {Canup}, R.~M., \& {Levison}, H.~F. 1999, \icarus, 142, 219,
  \dodoi{10.1006/icar.1999.6201}

\bibitem[{{Armstrong} {et~al.}(2019){Armstrong}, {Frost}, {McCammon}, {Rubie},
  \& {Boffa Ballaran}}]{2019SciArmstrong}
{Armstrong}, K., {Frost}, D.~J., {McCammon}, C.~A., {Rubie}, D.~C., \& {Boffa
  Ballaran}, T. 2019, Science, 365, 903, \dodoi{10.1126/science.aax8376}

\bibitem[{{Asphaug}(2010)}]{2010ChEGAsphaug}
{Asphaug}, E. 2010, Chemie der Erde / Geochemistry, 70, 199,
  \dodoi{10.1016/j.chemer.2010.01.004}

\bibitem[{{Asphaug} {et~al.}(2006){Asphaug}, {Agnor}, \&
  {Williams}}]{2006NatureAsphaug}
{Asphaug}, E., {Agnor}, C.~B., \& {Williams}, Q. 2006, \nat, 439, 155,
  \dodoi{10.1038/nature04311}

\bibitem[{{Asphaug} \& {Reufer}(2014)}]{2014NatGeoAsphaug}
{Asphaug}, E., \& {Reufer}, A. 2014, \natgeo, 7, 564, \dodoi{10.1038/ngeo2189}

\bibitem[{{Bonsor} {et~al.}(2015){Bonsor}, {Leinhardt}, {Carter}, {Elliott},
  {Walter}, \& {Stewart}}]{2015IcarBonsor}
{Bonsor}, A., {Leinhardt}, Z.~M., {Carter}, P.~J., {et~al.} 2015, \icarus, 247,
  291, \dodoi{10.1016/j.icarus.2014.10.019}

\bibitem[{{Burger} {et~al.}(2020){Burger}, {Bazs{\'o}}, \&
  {Sch{\"a}fer}}]{2020A&ABurger}
{Burger}, C., {Bazs{\'o}}, {\'A}., \& {Sch{\"a}fer}, C.~M. 2020, \aap, 634,
  A76, \dodoi{10.1051/0004-6361/201936366}

\bibitem[{{Cambioni} {et~al.}(2019){Cambioni}, {Asphaug}, {Emsenhuber},
  {Gabriel}, {Furfaro}, \& {Schwartz}}]{2019ApJCambioni}
{Cambioni}, S., {Asphaug}, E., {Emsenhuber}, A., {et~al.} 2019, \apj, 875, 40,
  \dodoi{10.3847/1538-4357/ab0e8a}

\bibitem[{{Carter} {et~al.}(2015){Carter}, {Leinhardt}, {Elliott}, {Walter}, \&
  {Stewart}}]{Carter2015apj}
{Carter}, P.~J., {Leinhardt}, Z.~M., {Elliott}, T., {Walter}, M.~J., \&
  {Stewart}, S.~T. 2015, \apj, 813, 72, \dodoi{10.1088/0004-637X/813/1/72}

\bibitem[{{Chambers}(1999)}]{1999MNRASChambers}
{Chambers}, J.~E. 1999, \mnras, 304, 793,
  \dodoi{10.1046/j.1365-8711.1999.02379.x}

\bibitem[{{Chambers}(2001)}]{2001IcarusChambers}
---. 2001, \icarus, 152, 205, \dodoi{10.1006/icar.2001.6639}

\bibitem[{{Chambers}(2013)}]{2013IcarusChambers}
---. 2013, \icarus, 224, 43, \dodoi{10.1016/j.icarus.2013.02.015}

\bibitem[{{Clement} {et~al.}(2019){Clement}, {Raymond}, \&
  {Kaib}}]{2019AJClementA}
{Clement}, M.~S., {Raymond}, S.~N., \& {Kaib}, N.~A. 2019, \aj, 157, 38,
  \dodoi{10.3847/1538-3881/aaf21e}

\bibitem[{{Dahl} \& {Stevenson}(2010)}]{2010EPSLDahlStevenson}
{Dahl}, T.~W., \& {Stevenson}, D.~J. 2010, Earth and Planetary Science Letters,
  295, 177, \dodoi{10.1016/j.epsl.2010.03.038}

\bibitem[{{de Vries} {et~al.}(2016){de Vries}, {Nimmo}, {Melosh}, {Jacobson},
  {Morbidelli}, \& {Rubie}}]{2016PEPSdeVries}
{de Vries}, J., {Nimmo}, F., {Melosh}, H.~J., {et~al.} 2016, Progress in Earth
  and Planetary Science, 3, 7, \dodoi{10.1186/s40645-016-0083-8}

\bibitem[{{Deguen} {et~al.}(2011){Deguen}, {Olson}, \& {Cardin}}]{2011Deguen}
{Deguen}, R., {Olson}, P., \& {Cardin}, P. 2011, \epsl, 310, 303,
  \dodoi{10.1016/j.epsl.2011.08.041}

\bibitem[{{Dwyer} {et~al.}(2015){Dwyer}, {Nimmo}, \&
  {Chambers}}]{2015IcarDwyer}
{Dwyer}, C.~A., {Nimmo}, F., \& {Chambers}, J.~E. 2015, \icarus, 245, 145,
  \dodoi{10.1016/j.icarus.2014.09.010}

\bibitem[{{Emsenhuber} \& {Asphaug}(2019)}]{2019ApJEmsenhuberA}
{Emsenhuber}, A., \& {Asphaug}, E. 2019, \apj, 875, 95,
  \dodoi{10.3847/1538-4357/ab0c1d}

\bibitem[{Emsenhuber \& Cambioni(2019)}]{2019SoftwareEmsenhuberCambioni}
Emsenhuber, A., \& Cambioni, S. 2019, collresolve, 1.1,  Zenodo,
  \dodoi{10.5281/zenodo.3560892}

\bibitem[{{Emsenhuber} {et~al.}(2020){Emsenhuber}, {Cambioni}, {Asphaug},
  {Gabriel}, {Schwartz}, \& {Furfaro}}]{EmsenhuberApj2020}
{Emsenhuber}, A., {Cambioni}, S., {Asphaug}, E., {et~al.} 2020, \apj, 691, 6,
  \dodoi{10.3847/1538-4357/ab6de5}

\bibitem[{{Fischer} {et~al.}(2017){Fischer}, {Campbell}, \&
  {Ciesla}}]{2017EPSLFischer}
{Fischer}, R.~A., {Campbell}, A.~J., \& {Ciesla}, F.~J. 2017, Earth and
  Planetary Science Letters, 458, 252, \dodoi{10.1016/j.epsl.2016.10.025}

\bibitem[{{Frost} {et~al.}(2010){Frost}, {Asahara}, {Rubie}, {Miyajima},
  {Dubrovinsky}, {Holzapfel}, {Ohtani}, {Miyahara}, \& {Sakai}}]{2010Frost}
{Frost}, D.~J., {Asahara}, Y., {Rubie}, D.~C., {et~al.} 2010, Journal of
  Geophysical Research (Solid Earth), 115, B02202, \dodoi{10.1029/2009JB006302}

\bibitem[{{Gabriel} {et~al.}(2020){Gabriel}, {Jackson}, {Asphaug}, {Reufer},
  {Jutzi}, \& {Benz}}]{2020ApJGabriel}
{Gabriel}, T. S.~J., {Jackson}, A.~P., {Asphaug}, E., {et~al.} 2020, \apj, 892,
  40, \dodoi{10.3847/1538-4357/ab528d}

\bibitem[{{Genda} {et~al.}(2017){Genda}, {Fujita}, {Kobayashi}, {Tanaka},
  {Suetsugu}, \& {Abe}}]{2017IcarusGenda}
{Genda}, H., {Fujita}, T., {Kobayashi}, H., {et~al.} 2017, \icarus, 294, 234,
  \dodoi{10.1016/j.icarus.2017.03.009}

\bibitem[{{Gessmann} {et~al.}(1999){Gessmann}, {Rubie}, \&
  {McCammon}}]{1999GeCoAGessmann}
{Gessmann}, C.~K., {Rubie}, D.~C., \& {McCammon}, C.~A. 1999, \gca, 63, 1853,
  \dodoi{10.1016/S0016-7037(99)00059-9}

\bibitem[{Girosi {et~al.}(1995)Girosi, Jones, \&
  Poggio}]{girosi1995regularization}
Girosi, F., Jones, M., \& Poggio, T. 1995, Neural computation, 7, 219

\bibitem[{Hagan {et~al.}(1997)Hagan, Demuth, \& Beale}]{demuth2014neural}
Hagan, M.~T., Demuth, H.~B., \& Beale, M. 1997, Neural network design (PWS
  Publishing Co.)

\bibitem[{{Hauck} {et~al.}(2013){Hauck}, {Margot}, {Solomon}, {Phillips},
  {Johnson}, {Lemoine}, {Mazarico}, {McCoy}, {Padovan}, \& {Peale}}]{2013Hauck}
{Hauck}, S.~A., {Margot}, J.-L., {Solomon}, S.~C., {et~al.} 2013, Journal of
  Geophysical Research (Planets), 118, 1204, \dodoi{10.1002/jgre.20091}

\bibitem[{{Helffrich}(2017)}]{2017HelffrichMarsCore}
{Helffrich}, G. 2017, Progress in Earth and Planetary Science, 4, 24,
  \dodoi{10.1186/s40645-017-0139-4}

\bibitem[{Holsapple(1994)}]{HOLSAPPLE19941067}
Holsapple, K. 1994, Planetary and Space Science, 42, 1067,
  \dodoi{https://doi.org/10.1016/0032-0633(94)90007-8}

\bibitem[{{Holsapple} \& {Housen}(1986)}]{1986MmSAI..57...65H}
{Holsapple}, K.~A., \& {Housen}, K.~R. 1986, \memsai, 57, 65

\bibitem[{Housen \& Holsapple(1990)}]{HOUSEN1990226}
Housen, K.~R., \& Holsapple, K.~A. 1990, Icarus, 84, 226,
  \dodoi{https://doi.org/10.1016/0019-1035(90)90168-9}

\bibitem[{{Jacobson} {et~al.}(2017){Jacobson}, {Rubie}, {Hernlund},
  {Morbidelli}, \& {Nakajima}}]{2017JacobsonEPSL}
{Jacobson}, S.~A., {Rubie}, D.~C., {Hernlund}, J., {Morbidelli}, A., \&
  {Nakajima}, M. 2017, \epsl, 474, 375, \dodoi{10.1016/j.epsl.2017.06.023}

\bibitem[{{Kegler} {et~al.}(2008){Kegler}, {Holzheid}, {Frost}, {Rubie},
  {Dohmen}, \& {Palme}}]{2008EPSLKegler}
{Kegler}, P., {Holzheid}, A., {Frost}, D.~J., {et~al.} 2008, Earth and
  Planetary Science Letters, 268, 28, \dodoi{10.1016/j.epsl.2007.12.020}

\bibitem[{{Kobayashi} {et~al.}(2019){Kobayashi}, {Isoya}, \&
  {Sato}}]{2019ApJKobayashi}
{Kobayashi}, H., {Isoya}, K., \& {Sato}, Y. 2019, \apj, 887, 226,
  \dodoi{10.3847/1538-4357/ab5307}

\bibitem[{{Kokubo} \& {Genda}(2010)}]{2010ApJKokubo}
{Kokubo}, E., \& {Genda}, H. 2010, \apjl, 714, L21,
  \dodoi{10.1088/2041-8205/714/1/L21}

\bibitem[{{Leinhardt} \& {Stewart}(2012)}]{2012ApJLeinhardt}
{Leinhardt}, Z.~M., \& {Stewart}, S.~T. 2012, \apj, 745, 79,
  \dodoi{10.1088/0004-637X/745/1/79}

\bibitem[{{Lock} \& {Stewart}(2019)}]{2019SciLockStewart}
{Lock}, S.~J., \& {Stewart}, S.~T. 2019, Science Advances, 5, eaav3746,
  \dodoi{10.1126/sciadv.aav3746}

\bibitem[{{Mann} {et~al.}(2009){Mann}, {Frost}, \& {Rubie}}]{2009GeCoAMann}
{Mann}, U., {Frost}, D.~J., \& {Rubie}, D.~C. 2009, \gca, 73, 7360,
  \dodoi{10.1016/j.gca.2009.08.006}

\bibitem[{{Melosh}(2007)}]{2007M&PSMelosh}
{Melosh}, H.~J. 2007, \maps, 42, 2079,
  \dodoi{10.1111/j.1945-5100.2007.tb01009.x}

\bibitem[{{Monaghan}(1992)}]{1992ARA&AMonaghan}
{Monaghan}, J.~J. 1992, \araa, 30, 543,
  \dodoi{10.1146/annurev.aa.30.090192.002551}

\bibitem[{{Nakajima} \& {Stevenson}(2016)}]{2016LPINakajima}
{Nakajima}, M., \& {Stevenson}, D.~J. 2016, in Lunar and Planetary Science
  Conference, Lunar and Planetary Science Conference, 2053

\bibitem[{{O'Brien} {et~al.}(2006){O'Brien}, {Morbidelli}, \&
  {Levison}}]{2006IcarusOBrien}
{O'Brien}, D.~P., {Morbidelli}, A., \& {Levison}, H.~F. 2006, \icarus, 184, 39,
  \dodoi{10.1016/j.icarus.2006.04.005}

\bibitem[{{Palme} \& {O'Neill}(2003)}]{2003PalmeNeill}
{Palme}, H., \& {O'Neill}, H. S.~C. 2003, Treatise on Geochemistry, 2, 568,
  \dodoi{10.1016/B0-08-043751-6/02177-0}

\bibitem[{{Quintana} {et~al.}(2016){Quintana}, {Barclay}, {Borucki}, {Rowe}, \&
  {Chambers}}]{2016ApJQuintana}
{Quintana}, E.~V., {Barclay}, T., {Borucki}, W.~J., {Rowe}, J.~F., \&
  {Chambers}, J.~E. 2016, \apj, 821, 126, \dodoi{10.3847/0004-637X/821/2/126}

\bibitem[{Raskin \& Morello(2019)}]{raskin2019material}
Raskin, C., \& Morello, C. 2019, EPSC, 2019, EPSC

\bibitem[{{Raymond} {et~al.}(2006){Raymond}, {Quinn}, \&
  {Lunine}}]{2006IcarusRaymond}
{Raymond}, S.~N., {Quinn}, T., \& {Lunine}, J.~I. 2006, \icarus, 183, 265,
  \dodoi{10.1016/j.icarus.2006.03.011}

\bibitem[{{Reufer}(2011)}]{2011PhDReufer}
{Reufer}, A. 2011, PhD thesis, University of Bern

\bibitem[{{Rosswog}(2009)}]{2009NARRosswog}
{Rosswog}, S. 2009, \nar, 53, 78, \dodoi{10.1016/j.newar.2009.08.007}

\bibitem[{Rubie {et~al.}(2015)Rubie, Nimmo, \& Melosh}]{RUBIE201543}
Rubie, D., Nimmo, F., \& Melosh, H. 2015, in Treatise on Geophysics (Second
  Edition), second edition edn., ed. G.~Schubert (Oxford: Elsevier), 43--79.
\newblock
  \url{https://www.sciencedirect.com/science/article/pii/B9780444538024001548}

\bibitem[{{Rubie} {et~al.}(2015){Rubie}, {Jacobson}, {Morbidelli}, {O'Brien},
  {Young}, {de Vries}, {Nimmo}, {Palme}, \& {Frost}}]{Rubie2015}
{Rubie}, D., {Jacobson}, S., {Morbidelli}, A., {et~al.} 2015, \icarus, 248, 89,
  \dodoi{10.1016/j.icarus.2014.10.015}

\bibitem[{{Rubie} {et~al.}(2016){Rubie}, {Laurenz}, {Jacobson}, {Morbidelli},
  {Palme}, {Vogel}, \& {Frost}}]{2016Rubie}
{Rubie}, D.~C., {Laurenz}, V., {Jacobson}, S.~A., {et~al.} 2016, Sci., 353,
  1141, \dodoi{10.1126/science.aaf6919}

\bibitem[{{Rubie} {et~al.}(2003){Rubie}, {Melosh}, {Reid}, {Liebske}, \&
  {Righter}}]{2003Rubie}
{Rubie}, D.~C., {Melosh}, H.~J., {Reid}, J.~E., {Liebske}, C., \& {Righter}, K.
  2003, \epsl, 205, 239, \dodoi{10.1016/S0012-821X(02)01044-0}

\bibitem[{{Rubie} {et~al.}(2011){Rubie}, {Frost}, {Mann}, {Asahara}, {Nimmo},
  {Tsuno}, {Kegler}, {Holzheid}, \& {Palme}}]{2011Rubie}
{Rubie}, D.~C., {Frost}, D.~J., {Mann}, U., {et~al.} 2011, \epsl, 301, 31,
  \dodoi{10.1016/j.epsl.2010.11.030}

\bibitem[{Sorokhtin {et~al.}(2011)Sorokhtin, Chilingar, \&
  Sorokhtin}]{2013EarthCMF}
Sorokhtin, O., Chilingar, G., \& Sorokhtin, N. 2011, in Developments in Earth
  and Environmental Sciences, Vol.~10, Evolution of Earth and its Climate:
  Birth, Life and Death of Earth, ed. O.~Sorokhtin, G.~Chilingarian, \&
  N.~Sorokhtin (Elsevier), 13 -- 60.
\newblock
  \url{http://www.sciencedirect.com/science/article/pii/B9780444537577000027}

\bibitem[{{Stewart} \& {Leinhardt}(2009)}]{2009ApJStewartLeinhardt}
{Stewart}, S.~T., \& {Leinhardt}, Z.~M. 2009, \apjl, 691, L133,
  \dodoi{10.1088/0004-637X/691/2/L133}

\bibitem[{{Stewart} \& {Leinhardt}(2012)}]{2012ApJ...751...32S}
---. 2012, \apj, 751, 32, \dodoi{10.1088/0004-637X/751/1/32}

\bibitem[{Thompson \& Lauson(1972)}]{ANEOS}
Thompson, S.~L., \& Lauson, H.~S. 1972, {Improvements in the CHART-D
  Radiation-hydrodynamic code III: Revised analytic equations of state}, Tech.
  Rep. SC-RR-71 0714, Sandia National Laboratories

\bibitem[{{Tonks} \& {Melosh}(1993)}]{1993TonksMelosh}
{Tonks}, W.~B., \& {Melosh}, H.~J. 1993, \jgr, 98, 5319,
  \dodoi{10.1029/92JE02726}

\bibitem[{{Turcotte} \& {Schubert}(2002)}]{2002geobook}
{Turcotte}, D.~L., \& {Schubert}, G. 2002, {Geodynamics - 2nd Edition}
  (Cambridge University Press), \dodoi{10.2277/0521661862}

\bibitem[{{Weidenschilling}(1977)}]{1977ApSSWeidenschilling}
{Weidenschilling}, S.~J. 1977, \apss, 51, 153, \dodoi{10.1007/BF00642464}

\bibitem[{{Wetherill}(1985)}]{1985ScienceWetherill}
{Wetherill}, G.~W. 1985, \sci, 228, 877, \dodoi{10.1126/science.228.4701.877}

\bibitem[{{Zahnle} {et~al.}(2007){Zahnle}, {Arndt}, {Cockell}, {Halliday},
  {Nisbet}, {Selsis}, \& {Sleep}}]{2007SSRZahnle}
{Zahnle}, K., {Arndt}, N., {Cockell}, C., {et~al.} 2007, \ssr, 129, 35,
  \dodoi{10.1007/s11214-007-9225-z}

\bibitem[{{Zahnle} {et~al.}(2020){Zahnle}, {Lupu}, {Catling}, \&
  {Wogan}}]{2020arXivZahnle}
{Zahnle}, K.~J., {Lupu}, R., {Catling}, D.~C., \& {Wogan}, N. 2020, The
  Planetary Science Journal, 1, 11, \dodoi{10.3847/PSJ/ab7e2c}

\bibitem[{{Zube} {et~al.}(2019){Zube}, {Nimmo}, {Fischer}, \&
  {Jacobson}}]{2019EPSLZube}
{Zube}, N.~G., {Nimmo}, F., {Fischer}, R.~A., \& {Jacobson}, S.~A. 2019, Earth
  and Planetary Science Letters, 522, 210, \dodoi{10.1016/j.epsl.2019.07.001}

\end{thebibliography}

\end{document}